\documentclass[prd,preprint,superscriptaddress,amsmath,amssymb,nofootinbib]{revtex4}

\usepackage{slashed,bbold}
\usepackage{graphicx,color}

\begin{document}

{\begin{flushright}{KIAS-P17037}
\end{flushright}}

\title{ Neutrino mass in a gauged $L_\mu - L_\tau$ model }
%
%
\author{Chuan-Hung Chen}
\email{physchen@mail.ncku.edu.tw}
\affiliation{Department of Physics, National Cheng-Kung University, Tainan 70101, Taiwan}

\author{Takaaki Nomura}
\email{nomura@kias.re.kr}
\affiliation{School of Physics, KIAS, Seoul 02455, Korea}

\date{\today}

\begin{abstract}
We study the origin of neutrino mass through  lepton-number violation and  spontaneous $U(1)_{L_\mu-L_\tau}$ symmetry breaking.  To accomplish the purpose,  we include one Higgs triplet, two singlet scalars, and two vector-like doublet leptons in the  $U(1)_{L_\mu-L_\tau}$ gauge extension of the standard model. To completely determine the free parameters, we employ the Frampton-Glashow-Marfatia (FGM) two-zero texture neutrino mass matrix as a theoretical input. It is found that  when some particular Yukawa couplings vanish, an FGM pattern can be achieved in the model. Besides the explanation of neutrino data, we  find that the absolute value of neutrino mass $m_j$ can be obtained in the model, and their sum can satisfy the upper bound of the cosmological measurement with $\sum_j |m_j| < 0.12$ eV. The effective Majorana neutrino mass for neutrinoless double-beta decay is below the current upper limit and is obtained as $\langle m_{\beta \beta} \rangle =(0.34,\, 2.3)\times 10^{-2}$ eV. In addition, the doubly charged Higgs $H^{\pm\pm}$ decaying to $\mu^\pm  \tau^\pm$ final states can be induced from a dimension-6 operator and is not suppressed, and its branching ratio is compatible with the $H^{\pm \pm}\to W^\pm W^\pm$ decay when the vacuum expectation value of Higgs triplet is $O(0.01)$ GeV.  
\end{abstract}

\maketitle

\section{Introduction}

In spite of the mass hierarchy among the quarks and charged leptons, the particle masses, with the exception of  the neutrinos, in the standard model (SM) can be attributed to the  Brout-Englert-Higgs (BEH) mechanism~\cite{Englert:1964et,Higgs:1964pj}, where the predicted  Higgs boson is observed using ATLAS~\cite{:2012gk} and CMS~\cite{:2012gu} at a mass of 125 GeV. Based on the neutrino oscillation experiments, it is found that the neutrinos are also massive particles; however, the definite origin of their masses  so far is unknown. 

Moreover, although nonzero neutrino masses have been determined by the experiments, we still cannot tell their mass order, i.e.,  $|m_1|< |m_2| < |m_3|$ or $|m_3|< |m_1| < |m_2|$ is possible, where the former and latter  are the mass spectrum with normal ordering (NO) and inverted ordering (IO), respectively. Hence, the current neutrino data can be shown in terms of the different mass ordering as~\cite{PDG}:
 \begin{align}
 \Delta m^2_{21} & = (7.53 \pm 0.18)\times 10^{-5} ~{\rm eV^2}\,, \ \sin^2\theta_{12} =0.304 \pm 0.014\,, \nonumber \\
 \Delta m^2_{32} & = (2.44 \pm 0.06,\, 2.51 \pm 0.06)\times 10^{-3}~ {\rm eV^2~(NO,\,IO) }\,, \nonumber \\
 \sin^2\theta_{23}&= (0.51 \pm 0.05,\, 0.50 \pm 0.05) ~ {\rm (NO,\, IO)} \,, \nonumber \\
 \sin^2\theta_{13} & = (2.19 \pm 0.12)\times 10^{-2}\,, \label{eq:nu_data}
 \end{align}
where $m^2_{21} \equiv m^2_2 -m^2_1$, $m^2_{23}$ denotes  $m^2_{3} -m^2_{2}$ for NO or $m^2_{2} -m^2_3$ for IO, and $\theta_{ij}$ are the mixing angles of  the Pontecorvo-Maki-Nakagawa-Sakata (PMNS) matrix~\cite{Pontecorvo:1957cp,Maki:1962mu}.  From the results, it is clearly seen that the PMNS matrix pattern is different from the Cabibbo-Kobayashi-Maskawa (CKM) matrix~\cite{Cabibbo:1963yz,Kobayashi:1973fv}, which dictates   the quark-flavor mixing. In this work,   we plan to study a model, where  based on a flavor symmetry, the neutrino masses are dynamically generated  without introducing singlet right-handed neutrinos~\cite{SeeSaw}, and  all neutrino data  can be explained.  Although it is inevitable to fine-tune the Yukawa couplings  to fit  the neutrino masses,  the model can  provide interesting phenomenological  implications in  flavor and collider physics. 
  
  Inspired by  the experimental indication of maximal  $\theta_{23}$, large $\theta_{12}$,  and small $\theta_{13}$, various Abelian flavor-symmetry based models have been proposed to understand the neutrino properties~\cite{Binetruy:1996cs,Mohapatra:1999zr,Grimus:2000kv,Bell:2000vh,Babu:2002ex,Goh:2002nk,Petcov:2004rk,Grimus:2004cj,Choubey:2004hn,Adhikary:2006rf,Asai:2017ryy}.  Among these flavor symmetries, we investigate the neutrino problems in an $U(1)_{L_\mu - L_\tau}$ gauge symmetry. We focus on such gauge symmetry  based on some phenomenological considerations: (i) gauge anomaly-free conditions are automatically satisfied~\cite{He:1990pn,Foot:1994vd}; (ii)  excess  of muon anomalous magnetic dipole moment (muon $g-2$) can be resolved~\cite{Gninenko:2001hx,Gninenko:2014pea,Altmannshofer:2016brv}; (iii)  excesses in semileptonic $B$-meson decays can be explained~\cite{Altmannshofer:2014cfa,Crivellin:2015mga,Altmannshofer:2016jzy,Ko:2017yrd,Chen:2017usq}; (iv) potential signals for the processes $e^+ e^- \to \gamma Z'$~\cite{Kaneta:2016uyt,Araki:2017wyg} and $\tau\to \mu Z'Z'$~\cite{Chen:2017cic} can be observed at Belle II. Other interesting studies can be found in~\cite{Heeck:2014qea,Baek:2015mna,Baek:2015fea,Heeck:2016xkh,Altmannshofer:2016oaq,Patra:2016shz,Biswas:2016yan,Lee:2017ekw,Heeck:2011wj}.
  
   In order to dynamically generate the neutrino masses, we require that each Majorana matrix entry  is  related to the lepton-number violating effect and  the breaking of spontaneous $U(1)_{L_\mu - L_\tau}$  symmetry.  To achieve the  lepton-number violation,  like type-II seesaw model~\cite{Magg:1980ut,Konetschny:1977bn}, we introduce a Higgs triplet, which carries a hypercharge $Y=1$ and has no $U(1)_{L_\mu-L_\tau}$ charge.  We find that due to the protection of $U(1)_{L_\mu -L_\tau}$ gauge symmetry, we cannot obtain a realistic Majorana neutrino mass matrix without further introducing  the  breaking of $U(1)_{L_\mu -L_\tau}$. Therefore, to break the  gauge symmetry, we employ two singlet scalars, which carry  different $U(1)_{L_\mu -L_\tau}$ charges.   Because the lepton chirality cannot be matched, the SM leptons cannot  couple to the singlet scalars; therefore, we must introduce proper exotic heavy leptons as the media. To avoid gauge anomalies, we employ two vector-like doublet leptons as the candidates.   Based on  the $U(1)_{L_\mu -L_\tau}$ gauge symmetry,  the number of singlet scalars and vector-like leptons (VLLs) in this approach is the minimal requirement by which to obtain a proper Majorana neutrino mass matrix.

  Since the number of free parameters in the Majorana neutrino mass matrix is more than that of  the neutrino data, not all free parameters can be determined. In order to completely determine the free parameters, we employ the Frampton-Glashow-Marfatia (FGM) matrix pattern~\cite{Frampton:2002yf}, which has two independent zeros, as a theoretical input.   
   
 It is demonstrated later that not all Yukawa couplings appearing in the neutrino mass matrix are small. Therefore,  in addition to the neutrino issue,  the model can also provide  interesting phenomena related to flavor and collider physics. For instance,  the lepton-flavor violating $h\to \mu \tau$ decay can be as large as the current measurements~\cite{Dorsner:2015mja,Herrero-Garcia:2016uab}; excess of muon $g-2$ can be resolved by the mediation of the $Z'$ gauge boson, and the doubly charged Higgs decaying to $\mu \tau$ and $WW$ can be compatible each other without requiring the VEV of the Higgs triplet to be the eV.  
 
 This paper is organized as follows. In Sec.~II, we introduce the model under the $SU(2)_L\times U(1)_Y\times U(1)_{L_\mu-L_\tau}$ local gauge symmetry. In Sec.~III, we generate the Majorana neutrino mass matrix without right-handed neutrinos in the model and discuss the relation to the FGM matrix pattern. The numerical analysis on neutrino physics and implications of the model on other phenomena are shown in Sec.~IV. A summary is given in Sec.~IV.

\section{Model}

 In this section, we  introduce the model under the $SU(2)_L\times U(1)_Y\times U(1)_{L_\mu-L_\tau}$ local gauge symmetry. In order to dynamically generate the neutrino mass in the $U(1)_{L_\mu-L_\tau}$ extension of the SM, in addition to the SM particles, we include one Higgs triplet ($\Delta$), two  vector-like doublet leptons ($L_4,\, L_5$), and two singlet scalars ($S, S'$). Their $U(1)_{L_\mu-L_\tau}$ charges are given in Table~\ref{tab:U1}, where the SM particles not shown in the table carry no such $U(1)$ charges. %
\begin{table}[hptb]
\caption{$U(1)_{L_\mu-L_\tau}$ charges of involving leptons, $S$, and $S'$.}
\begin{tabular}{ccccccccc} \hline \hline
            &  ~~~$e$~~~  & ~~~$\mu$~~~ & ~~~ $\tau$ ~~~& ~~~ $L_4$  ~~~ & ~~~ $L_5$  ~~~& ~~~$\Delta$~~~& ~~~$S'$~~~& ~~~$S$~~~\\ \hline 
 $U(1)$ &   0   &   1      & $-1$      &  $-1$  & $1$ & $0$ & $1$ & $2$\\ \hline \hline

\end{tabular}
\label{tab:U1}
\end{table}
 Accordingly,  the  Yukawa couplings to the Higgs triplet  are written as:
 \begin{align}
 -{\cal L}^{\Delta}_Y & =  \frac{1}{2} Y_{ee}  L^T_e C i\tau_2 \Delta L_e + Y_{\mu\tau} L^T_\mu C i \tau_2 \Delta L_\tau   +Y_{\mu 4} L^T_\mu C i\tau_2 \Delta L_{4L} \nonumber \\
 & + Y_{\tau 5} L^T_\tau C i\tau_2 \Delta L_{5L}  + Y_{45} L^T_{4L} C i\tau_2 \Delta L_{5L}   + Y'_{45} L^T_{4R} C i\tau_2 \Delta L_{5R} + H.c. \label{eq:Yukawa1}
 \end{align}
  From the above equation, if the Higgs triplet $\Delta$ carries  two  lepton-number units, the Yukawa interactions are lepton-number conserved. However, when the Higgs triplet obtains a VEV, i.e. $\langle \Delta \rangle =v_\Delta/\sqrt{2}$, the lepton-number violating Majorana neutrino mass matrix for three light neutrinos is induced and expressed as:
   \begin{align}
   M^{\nu} &=  \left(
\begin{array}{ccc}
\frac{ Y_{ee} v_\Delta}{\sqrt{2} }&  0 &   0\\
  0 & 0  & \frac{Y_{\mu\tau}v_\Delta}{\sqrt{2} } \\
0  &  \frac{Y_{\mu\tau}v_\Delta}{\sqrt{2}}  &   0
\end{array}
\right)\,, \label{eq:Mmass}
   \end{align}
 where the pattern of mass matrix leads to $m_2=m_3$,  $\theta_{13}=\theta_{12}=0$, and $\theta_{23}=\pi/4$~\cite{Binetruy:1996cs,Choubey:2004hn,Adhikary:2006rf}. Obviously, the results cannot explain the current neutrino data~\cite{PDG}.  We clearly demonstrate that the  neutrino mass matrix, which arises  from the breaking of the electroweak symmetry and lepton-number violations,  cannot explain the neutrino  data due to the protection of $U(1)_{L_\mu-L_\tau}$ gauge invariance. In order to obtain a realistic neutrino mass matrix, we have to rely on other pieces of Yukawa interactions, which can break the $U(1)$ symmetry.  Concerning the magnitude of $v_\Delta$, according to the electroweak symmetry breaking,  the electroweak $\rho$-parameter at the tree-level  can be written as~\cite{Konetschny:1977bn}:
   \begin{equation}
   \rho=\frac{m^2_W}{m^2_{Z} c^2_{\theta_W} }= \frac{1+2v^2_\Delta/v^2_H}{1+4 v^2_\Delta/v^2_H}\,.
   \end{equation}
 Taking  the current precision measurement for $\rho$-parameter within $2\sigma$ errors,  the VEV of $\Delta$ has to be less than $3.4$ GeV. 
 
\begin{figure}[hpbt]
  \begin{center}
\includegraphics[scale=0.55]{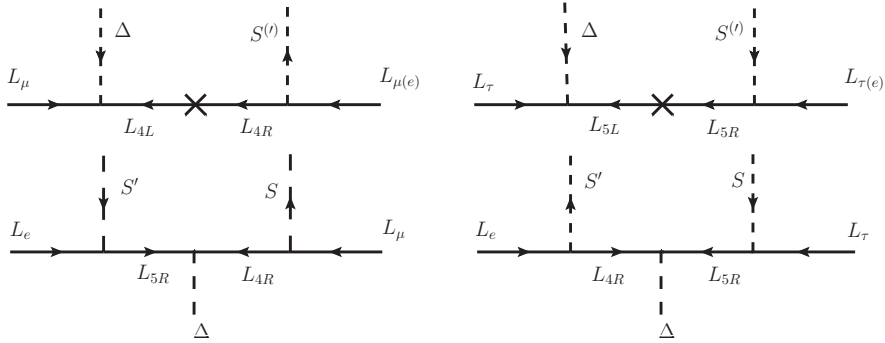}  
\caption{ Sketched Feynman diagrams for the Majorana neutrino mass matrix elements.}
\label{fig:one-loop}
\end{center}
\end{figure}

 In addition to  Eq.~(\ref{eq:Yukawa1}),  the gauge invariant Yukawa couplings to the Higgs and $S^{(\prime)}$ are given by:
 \begin{align}
 -{\cal L}_Y & = Y_{\ell} \bar L_{\ell} H \ell_R + y_\mu \bar L_{5L} H \mu_{ R} + y_\tau \bar L_{4L} H \tau_R  + y'_\mu \bar L_{\mu}  L_{4R} S +  y'_\tau \bar L_{\tau} L_{5R} S^\dagger   \nonumber \\
& + y_e \bar L_{e} L_{4R} S' + y'_e \bar L_{e} L_{5R} S'^\dagger  + y_S \bar L_{5L} L_{4R} S +y'_S   \bar L_{4L} L_{5R} S^\dagger  \nonumber \\
&  + m_{4L} \bar L_{4L} L_{4R}+ m_{5L}  \bar L_{5L} L_{5R} +   m_{4 \tau} \bar L_{4R} L_\tau + m_{5\mu} \bar L_{5R} L_\mu  + H.c.\,,   \label{eq:Yukawa2}
  \end{align} 
where $H$ is the SM Higgs doublet; only the first term is from the SM, and the other terms are the new Yukawa interactions. Although Eq.~(\ref{eq:Yukawa2}) can cause rich interesting phenomena for  lepton-flavor  physics,  we only focus on neutrino physics in this work, and a detailed study on the flavor physics can be found in~\cite{Chen:2017cic}. 
Based on the Yukawa interactions in Eq.~(\ref{eq:Yukawa2}), it is found that the new entries of the Majorana mass matrix can be induced from  higher dimensional operators, where the Feynman diagrams  are sketched in Fig.~\ref{fig:one-loop}, and the associated gauge invariant dimension-5 and -6 operators can be formulated as:
 \begin{align}
  -{\cal L}_{Y} & \supset  \frac{Y_{\mu4} y'^*_\mu}{m_{4L}} L^T_\mu C \bar \Delta L_\mu S^\dagger +  \frac{Y_{\tau 5} y'^*_\tau}{m_{5L}} L^T_\tau C \bar \Delta L_\tau S  
  + \left( \frac{Y_{\mu 4} y^*_e }{m_{4L}} + \frac{y^*_e Y'_{45} m_{5\mu}}{m_{4L} m_{5L}}  \right) L^T_e C \bar \Delta L_\mu S'^\dagger  \nonumber \\
  &  + \left( \frac{Y_{\tau 5} y'^*_e }{m_{5L}} + \frac{y'^*_e Y'_{45} m_{4\tau}}{m_{4L} m_{5L}}  \right) L^T_e C \bar \Delta L_\tau S' + \frac{Y'_{45} (y_e y'_e)^* }{m_{4L} m_{5L}} L^T_e C \bar\Delta L_e S' S'^\dagger   \nonumber \\
  & +  \frac{y'^*_{e}Y'_{45} y'^*_\mu}{m_{4L} m_{5L} } L^T_e C \bar \Delta L_\mu S^\dagger S' +  \frac{y^*_{e}Y'_{45} y'^*_\tau}{m_{4L} m_{5L} } L^T_e C \bar \Delta L_\tau S S'^\dagger + \frac{Y_{\mu 4} y'_S y'^*_\tau}{m_{4L} m_{5L}} L^T_\mu C \bar\Delta L_\tau S S^\dagger  \nonumber \\
  & + \frac{Y_{\tau 5} y_S y'^*_\mu}{m_{4L} m_{5L}} L^T_\tau C \bar\Delta L_\mu S S^\dagger + \frac{ (Y_{45}+Y'_{45}) y'^*_\tau y'^*_\mu }{m_{4L} m_{5L}} L^T_\mu C \bar \Delta L_\tau S S^\dagger \nonumber \\
&  + \frac{Y_{\mu 4} m_{4\tau}}{m_{4L}} L^T_\mu C \bar \Delta L_\tau +  \frac{Y_{\tau 5} m_{5\mu}}{m_{5L}} L^T_\tau C \bar \Delta L_\mu + H.c. \label{eq:lang}
 \end{align}
 with $\bar \Delta = i\tau_2 \Delta$. From the effective Lagrangian, when the $U(1)_{L_\mu-L_\tau}$ gauge symmetry is spontaneously broken by  $\langle S \rangle=v_{S}/\sqrt{2}$ and $\langle S' \rangle =v_{S'}/\sqrt{2}$,   the vanishing elements in Eq.~(\ref{eq:Mmass}) can be generated from Eq.~(\ref{eq:lang}) with $\langle \Delta \rangle = v_{\Delta}/\sqrt{2}$.  We note that the dimension-6 operator $L^T_\mu C \bar\Delta \bar \Delta^\dagger \bar \Delta L_\tau$ has been dropped due to $v_{\Delta}\ll v_{S,S'}$.   From Eq.~(\ref{eq:lang}), it can be seen that after electroweak symmetry breaking, the $m_{4\mu}$ and $m_{5\tau}$ effects can be combined with other terms as:
  \begin{align}
 Y_{\mu 4 } + \frac{m_{5\mu}}{m_{5L}} Y'_{45}  & \to \tilde Y_{\mu 4}   \,, \nonumber \\
  Y_{\tau 5} + \frac{m_{4\tau}}{m_{4L}} Y'_{45} & \to  \tilde Y_{\tau 5}  \,, \nonumber \\
 \frac{v^2_S }{2 m_{4L} m_{5L}} y'_S y'^*_\tau + \frac{m_{4\tau}}{m_{4L}} & \to    \frac{v^2_S }{2m_{4L} m_{5L}} \tilde y'_S y'^*_\tau \,, \nonumber \\
 \frac{v^2_S }{2 m_{4L} m_{5L}} y_S y'^*_\mu + \frac{m_{5\mu}}{m_{5L}} & \to   \frac{v^2_S }{2 m_{4L} m_{5L}} \tilde y_S y'^*_\mu  \,.
  \end{align}
Thus,  to fit the neutrino masses, we need to take $m_{4\tau, 5\mu}\ll m_{4L,5L}$.

 Since the neutrino masses are  generated by  the spontaneous $U(1)_{L_\mu-L_\tau}$ symmetry breaking,  we need to find the necessary conditions for vacuum stability. We thus write the gauge invariant scalar potential in this model as:
\begin{align}
{\cal V}=&
m_H^2 H^\dagger H +  m^2_\Delta {\rm Tr}[\Delta^\dagger \Delta] + m_{S'}^2 S'^\dagger S' + m_{S}^2 S^\dagger S  +\mu_\Delta [H^T (i\tau_2) \Delta^\dagger H +{\text{h.c.}}] \nonumber \\
& + \mu_S [S' S' S^\dagger + {\text{h.c.}}]
+ \lambda_{1} |H^\dag H|^2  + \lambda_{2} ({\rm Tr}[\Delta^\dagger \Delta])^2+ \lambda_{3} {\rm Tr}[(\Delta^\dagger \Delta)^2] + \lambda_{4} |S'^\dagger S'|^2 
 \nonumber \\
 &+ \lambda_{5} |S^\dagger S|^2 + \lambda_{6}(H^\dagger H) {\rm Tr}[\Delta^\dag\Delta] + H^\dagger \left( \lambda_7 \Delta \Delta^\dagger + \lambda_8 \Delta^\dagger \Delta \right) H + \lambda_{9} (S'^\dagger S') (H^\dagger H) \nonumber \\
 &+ \lambda_{10} (S^\dagger S) (H^\dagger H) 
+ \lambda_{11} (S'^\dagger S') {\rm Tr}[\Delta^\dagger \Delta] + \lambda_{12} (S^\dagger S) {\rm Tr}[\Delta^\dagger \Delta] + \lambda_{13} (S'^\dagger S')(S^\dagger S)\,. \label{eq:V}
 \end{align}
  The VEVs of scalar fields are obtained by the minimal conditions $\partial \langle \mathcal{V} \rangle/ \partial v_{H, S, S', \Delta} = 0$, and  each condition can be expressed as: 
\begin{align}
 \frac{\partial {\cal \langle V \rangle}}{\partial v_H} & \simeq m_H^2 v_H + \lambda_1 v_H^3 + \frac{1}{2} \lambda_{9} v_{S'}^2 v_H + \frac{1}{2} \lambda_{10} v_S^2 v_H \simeq 0, \\
\frac{\partial {\cal \langle V \rangle}}{\partial v_{S}} & \simeq m_{S}^2 v_{S} + \frac{1}{\sqrt{2}} \mu_S v_{S'}^2  +  \lambda_5 v_{S}^3  + \frac{1}{2} \lambda_{10} v_H^2 v_{S} + \frac{1}{2} \lambda_{13} v_{S'}^2 v_{S} \simeq 0, \\
\frac{\partial {\cal \langle V \rangle}}{\partial v_{S'}} & \simeq m_{S'}^2 v_{S'} + \sqrt{2} \mu_S v_{S'} v_S +  \lambda_4 v_{S'}^3  + \frac{1}{2} \lambda_{9} v_H^2 v_{S'}  + \frac{1}{2} \lambda_{13} v_{S}^2 v_{S'} \simeq 0, \\
\frac{\partial {\cal \langle V \rangle}}{\partial v_\Delta} & \simeq m_\Delta^2 v_\Delta + \frac{1}{\sqrt{2}} \mu_\Delta v_H^2 + \frac{1}{2} (\lambda_6 + \lambda_7) v_H^2 v_\Delta + \frac{1}{2} \lambda_{11} v_{S'}^2 v_\Delta + \frac{1}{2} \lambda_{12} v_S^2 v_\Delta \simeq 0, \label{eq:VEV_D}
\end{align}
where  we have ignored the $v_\Delta $ terms  in the first three equations and the $v^3_\Delta$ terms in the last equation  due to $v_\Delta \ll  v_{H,S,S'}$. In order to avoid the precision Higgs measurements, we can assume   the mixing between $H$ and $S(S')$ to be small,  where the scalar mixing is discussed below; then, the VEV of $H$ can be simplified as $v_{H} \approx \sqrt{- m^2_H/\lambda_1}$. If we further assume $\lambda_{13}$ and $\mu_S$ to be small, the VEVs of $S$ and $S'$ can be found as $v_S \approx \sqrt{- m^2_{S}/\lambda_5}$ and $v_{S'}\approx \sqrt{-m^2_{S'}/\lambda_4}$ with $m^2_{S,S'}<0$. The $v_{S}$ and $v_{S'}$ are free parameters and their relation to  the $Z'$-boson mass is given by $m^2_{Z'}=g^2_{Z'} ( 4 v^2_S + v^2_{S'})$;  hence,  their magnitudes can be taken as the  electroweak scale. From Eq.~(\ref{eq:VEV_D}),  the VEV of Higgs triplet can be determined as~\cite{Chen:2014lla}:
\begin{equation}
v_\Delta \simeq - \frac{1}{\sqrt{2}} \frac{\mu_\Delta v^2_H }{m_\Delta^2 + (\lambda_6+\lambda_7)v_H^2/2 + \lambda_{11} v_{S'}^2/2 + \lambda_{12} v_S^2/2}\,. \label{eq:vD}
\end{equation}
Because  $v_\Delta < 3.4$ GeV, in order to obtain the heavy Higgs triplet bosons, unlike the Higgs doublet and $S(S')$, $m^2_\Delta$ has to be positive and must also dictate the masses of the Higgs triplet bosons. From Eq.~(\ref{eq:vD}),  it can be seen that similar to the type-II seesaw model~\cite{Magg:1980ut,Konetschny:1977bn}, the Higgs triplet VEV is directly related to  the lepton-number soft breaking term. 

{We make a remark on the  oblique parameter constraint. For the Higgs triplet,  the mass difference between the  Higgs triplet components is predominantly dictated by the oblique $T$-parameter, where the mass splitting between singly and doubly charged Higgs mass is bounded  as $|m_{H^{\pm \pm}} - m_{H^\pm(H^0)}| \lesssim 50$ GeV~\cite{Chun:2012jw,Baak:2014ora,Das:2016bir}. Since our study does not directly related to the mass splitting of the Higgs triplet,  we can take $m_{H^{\pm \pm}} \approx m_{H^\pm(H^0)}$  to satisfy the constraint.  Similarly, because  the particle masses in  $L_{4(5)}$ are  taken to be the same, the vector-like leptons contributing to  the $T$-parameter are small  and can be neglected. }

{Although the involved new scalars do not directly affect the neutrino physics  in this study, it is of interest to understand the limit from the current SM Higgs precision measurements. Thus, we briefly discuss the mixings among the SM Higgs and new scalar bosons in the following analysis. Since the mixing between the SM Higgs and the Higgs triplet is suppressed by the small VEV of the Higgs triplet field, therefore, we consider the situations in the SM Higgs and singlet scalars. Moreover,  if we taking  $\lambda_{9,13} \ll 1$, the mixing between the SM Higgs and $S'$ is suppressed and can be neglected. Thus,  in order to show the constraint of the Higgs precision measurements, we only focus on the $H$-$S$ mixing. Using  the scalar potential in Eq.~(\ref{eq:V}) and  $H^T = (G^+, (v + \tilde h + i G^0)/\sqrt{2})$ and $S = (v_S + \tilde s + i \eta_S)/\sqrt{2}$, where $G^+$ and $G^0$ are the Nambu-Goldstone bosons in the SM, the squared mass matrix for the  $\tilde h$ and $\tilde s$  scalar bosons can be obtained as:
\begin{equation}
\mathcal{L} \supset \frac{1}{2} \begin{pmatrix} \tilde h \\ \tilde s \end{pmatrix}^T \begin{pmatrix} \lambda_1 v^2 & \frac{\lambda_{10}}{2} v v_S \\  \frac{\lambda_{10}}{2} v v_S  & \lambda_5 v_S^2 \end{pmatrix} \begin{pmatrix} \tilde h \\ \tilde s \end{pmatrix}.
\end{equation} 
Using the $2\times 2$ orthogonal matrix, written as:
\begin{equation}
\begin{pmatrix} h \\ s \end{pmatrix} = \begin{pmatrix} \cos \alpha & \sin \alpha \\ - \sin \alpha & \cos \alpha \end{pmatrix} \begin{pmatrix} \tilde h \\ \tilde s \end{pmatrix}\,,\label{eq:scalar-mass-fields}
\end{equation}
the eigenvalues of the mass-square matrix and the mixing angle $\alpha$ can be obtained as:
\begin{align}
m_{h, s}^2 &= \frac{\lambda_1 v^2 +\lambda_5 v_S^2 }{2} \pm \frac{1}{2} \sqrt{\left( \lambda_1 v^2 -\lambda_5 v_S^2 \right)^2 +  \lambda_{10}^2 v^2 v_S^2 }\,, \nonumber \\
\sin 2 \alpha &= \frac{ \lambda_{10} v v_S}{m_h^2 - m_S^2}\,, \label{eq:mixingS}
\end{align}
where $\alpha$ is the mixing angle, and $h$ is identified as the SM-like Higgs boson.
It is clearly seen that in addition to the VEV of the Higgs field, the mixing effect of $h$ and $s$ is associated with the $\lambda_{10}$ parameter and the VEV of $S$ field. 

 Although there are several channels for the SM-like Higgs production and decays, the most accurate  measurement in the LHC is the gluon-gluon fusion (ggF) Higgs production and the Higgs diphoton decay, i.e.  $pp(gg)\to h \to \gamma\gamma$. Thus, we only concentrate on the $h\to \gamma \gamma$ mode. 
For illustrating the influence of the new physics effects, we use the signal strength for $pp\to h\to \gamma\gamma$, defined as:
 \begin{align}
 \mu_{\gamma\gamma} = \frac{\sigma(pp\to h)_{\rm SM+NP}}{\sigma(pp\to h)_{\rm SM}} \frac{BR(h\to \gamma \gamma)_{\rm SM+ NP}}{BR(h\to \gamma \gamma)_{\rm SM}}\,, \label{eq:mu_2gamma}
 \end{align}
where the ATLAS and CMS results  using  luminosities of 80 fb$^{-1}$ and 35 fb$^{-1}$ at $\sqrt{s}=13$ TeV  are given by $\mu_{ggF}=0.97^{+0.15}_{-0.14}$~\cite{ATLAS:2018uso} and $\mu_{ggF}=1.02^{+0.19}_{-0.18}$~\cite{Sirunyan:2018ouh}, respectively. According to the current data,  we can take  $\delta\mu^{\rm NP}_{\gamma\gamma} =  \mu_{\gamma\gamma} -1 = \pm 15\%$ to  constrain the new physics effect. 

From Eq.~(\ref{eq:mixingS}), the SM Higgs couplings are modified by a factor of  $\cos \alpha$; thus, 
 we obtain $\sigma(pp\to h)_{\rm SM+NP} \simeq \cos^2 \alpha \times \sigma(pp\to h)_{\rm SM}$. For the $h$ decays, in addition to the SM channels, the $h$ can also decay into the $ss$ and $Z'Z'$ final states when kinematically  allowed in this model.
 In order to include these two decay modes, we write the relevant interactions as:
 \begin{align}
\label{eq:intH}
\mathcal{L} \supset  4 g_{Z'}^2 v_S \sin\alpha h Z'_\mu Z'^\mu - \frac{1}{2} g_{h s s} h s s\,,
\end{align}
where with $\lambda_1 \simeq (m_h/v)^2$ and $\lambda_5 \simeq (m_s/v_S)^2$, the effective coupling $g_{h s s}$ from the scalar potential  can be obtained  as:
\begin{align}
g_{h \phi\phi} & \simeq 6 \sin\alpha \cos\alpha \left( \frac{m_h^2}{v} \sin\alpha + \frac{m_s^2}{v_S} \cos\alpha \right) \nonumber \\
&+ \lambda_{10} (v \cos^3\alpha + v_S \sin^3\alpha  - 2 v_S \sin\alpha \cos^2 \alpha   - 2 v \sin^2 \alpha \cos \alpha)\,.
\end{align}
Accordingly, the partial decay rates for the $h \to ss$ and $h \to Z'Z'$ processes can be formulated as:
\begin{align}
\Gamma_{h \rightarrow Z' Z'} & = \frac{2{g'}^4 v_S^2 \sin^2 \alpha}{ \pi m_h} \sqrt{1 - \frac{4 m_{Z'}^2}{m_{h}^2}}
\left( 2 + \frac{m_{h}^4}{4 m_{Z'}^4} \left( 1 - \frac{2 m_{Z'}^2 }{m_{h}^2} \right)^2 \right)\,,  \nonumber \\
\Gamma_{h \rightarrow s s} &=\frac{g_{h s s}^2}{32\pi m_{h}} \sqrt{1 - \left(\frac{2m_s}{m_{h} } \right)^2}\,.
\end{align}
As a result, the $\mu_{\gamma \gamma}$ signal strength in Eq.~(\ref{eq:mu_2gamma}) can be obtained as: 
\begin{equation}
\mu_{\gamma \gamma} = \cos^4 \alpha \frac{\Gamma_h^{SM}}{\cos^2 \alpha \Gamma_h^{SM} + \Gamma_{h \to ss} + \Gamma_{h \to Z' Z'}}\,,
\end{equation}
where $\Gamma_h^{SM} \simeq 4.07$ MeV is the decay width of the SM Higgs~\cite{Dittmaier:2011ti}. {Using Eq.~(\ref{eq:mixingS}) and $v_S = m_{Z'}/(\sqrt{5} g_{Z'})$, which arises from $v_{S}=v_{S'}$,  we show $\delta\mu^{\rm NP}_{\gamma\gamma}$ as a function of $\lambda_{10}$ in the left panel of Fig.~\ref{fig:diphoton},  where $m_{Z'} = 0.2$ GeV and $g_{Z'} = 10^{-3}$ motivated from the muon $g-2$ are used.  With  $m_s = 10(200)$ GeV, the upper limit of $\lambda_{10}$ can be $\sim 0.01 (0.05)$, whereas the corresponding value of $\sin\alpha$ is $\sim 0.004(0.01)$.  Since we focus on a light $S$-boson in the phenomenological analysis, the effects of the  small mixing $\alpha$ angle can be  neglected.  In the considered  parameter region,  $s$ and $Z'$ mainly decay into $Z'Z'$ and $\bar \nu \nu$, respectively, it is of interest to see the constraint from the invisible Higgs decays, where the current upper limit of  branching ratio (BR) is $BR(h\to {\rm  invisible})< 0.24$~\cite{Aad:2015pla,Khachatryan:2016whc}. Thus, we show $BR(h \to ss)$ (dotted), $BR(h \to Z'Z')$ (dashed), and  $BR(h \to ss+ Z'Z')$ (solid) as a function of $\lambda_{10}$ in the right panel of Fig.~\ref{fig:diphoton}. It can be clearly seen  that the constraint from $\delta\mu^{\rm NP}_{\gamma\gamma}$ is  stricter than that from the  invisible Higgs decays. 
     
}

  \begin{figure}[hpbt]
  \begin{center}
\includegraphics[width=80mm]{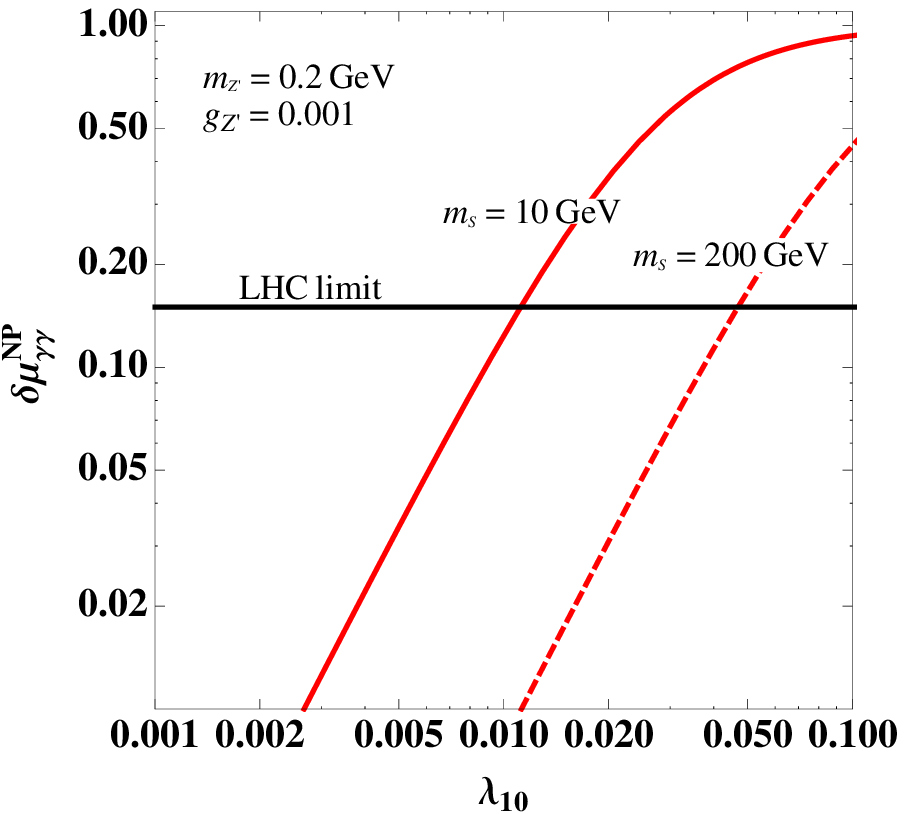}  
\includegraphics[width=80mm]{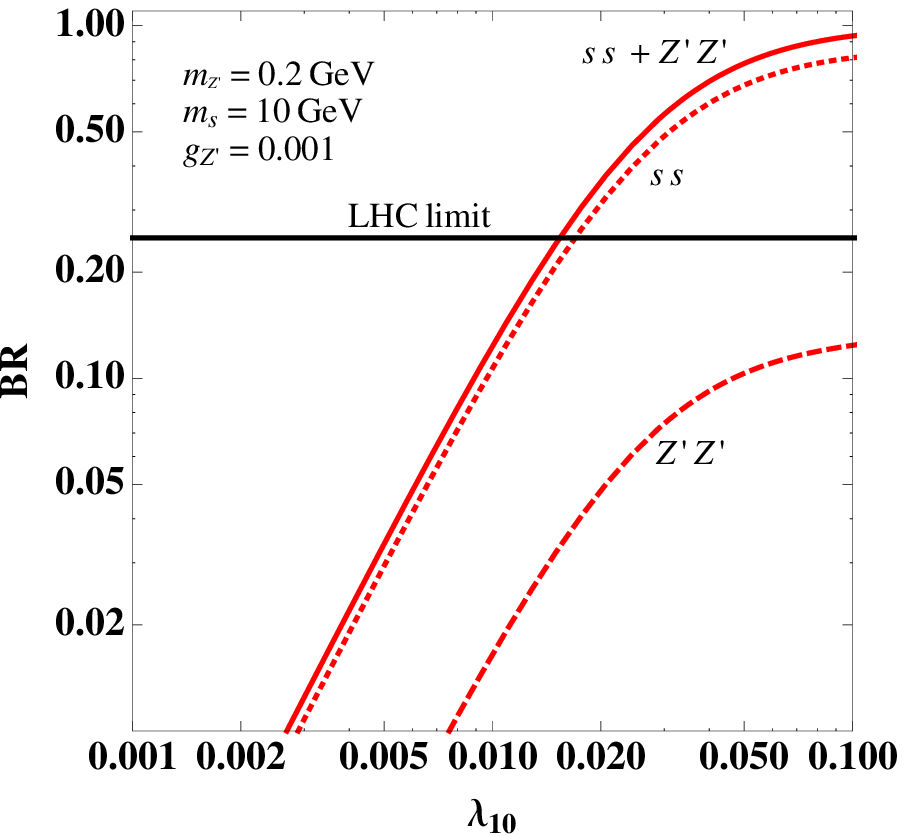}  
\caption{ Left: $\delta\mu^{\rm NP}_{\gamma\gamma}$ as a function of $\lambda_{10}$ with $m_s=200$ GeV (dashed) and $m_s=20$ GeV (solid), where the taken values of the other parameters are shown on the plot. Right: BRs for  the $h \to ss$ (dotted), $h \to Z'Z'$ (dashed), and $h\to ss+Z'Z'$ (solid) decays. The horizontal line denotes the experimental upper bound. }
\label{fig:diphoton}
\end{center}
\end{figure}

\section{ Charged lepton flavor mixing matrix}

 Since the PMNS matrix is related to the neutrino and charged-lepton flavor mixing matrices,  before discussing the neutrino mass generation in this model, we first analyze the possibly sizable charged-lepton flavor mixing. As mentioned before, the active neutrino mass matrix is dictated by the Yukawa couplings in Eqs. (\ref{eq:Yukawa1}) and (\ref{eq:Yukawa2}), therefore,  to explain the neutrino masses below the eV scale, most parameters have to be many orders  of magnitude smaller than one. On the other hand, in order to have  implications on the flavor physics, such as $h \to \mu \tau$ and $H^{--} \to \mu \tau$, we need  some parameters  to be of $O(10^{-2}-10^{-1})$. In order to simplify the analysis on the charged-lepton flavor mixing, we thus ignore the small parameters, which are dictated by the neutrino masses,  select the potentially sizable parameters, such as $y_{\mu,\tau}$, $y'_{\mu,\tau}$, and $Y_{45}$, and use these parameters to formulate the  flavor mixing matrix. The reason to select these parameters will be clear in the later analysis.

The SM charged leptons and the introduced heavy leptons   form a multiplet state in flavor space, denoted by $ \ell'^T= (\pmb{\ell\,, \Psi}_\ell )$ with $\pmb{\ell}=(e, \mu,\tau)$ and $\pmb{\Psi}^T_{\ell}=( L_4, L_5)$. From Eq.~(\ref{eq:Yukawa2}), the $5\times 5$ lepton mass matrix can be written as:
  \begin{equation}
  \bar\ell'_L  M_{\ell'}  \ell'_R =  \left(\begin{array}{cc}
 \pmb{ \bar \ell}_L \,, & \pmb{\bar \Psi}_{\ell L}  \end{array} \right) 
    \left( \begin{array}{c|c}
  ~~~{\pmb{m}_\ell}_{3\times 3}~~~ &  \pmb{ \delta m}_1  \\ \hline
 \pmb{\delta m}^T_2  &  \pmb{m}_L
 \end{array} \right)_{5\times 5} 
  \left(\begin{array}{c}
   \pmb{\ell}_R \\
   \pmb{\Psi}_{\ell R} \end{array}\right) \,, \label{eq:mass}
  \end{equation}
  where diag$\pmb{m_\ell}=( m_e, m_\mu, m_\tau)$, $m_f = v_H Y_f/\sqrt{2}$, diag$\pmb{m}_L = (m_{4L}, m_{5L})$,  and $\pmb{\delta m}_{1,2}$  are given by:
  \begin{align}
  \pmb{\delta m}^T_1 & = \left(\begin{array}{ccc}
 0 \,, & \frac{v_S y'_\mu }{\sqrt{2}}\,, &  0 \\
  0\,, &   0\,, & \frac{v_S y'_\tau}{\sqrt{2}}
 \end{array} \right)  \,, ~~
  \pmb{\delta m}^T_2 = \left(\begin{array}{ccc}
 0 \,, & 0\,, &  \frac{v_H y_\tau }{\sqrt{2}} \\
  0\,, & \frac{v_H y_\mu }{\sqrt{2}} \,, & 0
 \end{array} \right)\,.
     \end{align}
 The mass matrix $M_{\ell'}$ in Eq.~(\ref{eq:mass}) can be diagonalized by the  unitary matrices $U^{R, L}_\ell$ through $M^{\rm dia}_{\ell'} =  U^L_\ell M_{\ell'} U^{R^\dagger}_{\ell}$. Due to $v_{H,S}\ll m_{4L,5L}$, we can expand the flavor mixing effects in terms of  $v_{H,S}/m_{4L,5L}$; therefore,  the  $5\times 5$ flvaor mixing matrices  can be simplified as:
     \begin{align}
 U^\chi_\ell \approx  \left( \begin{array}{c|c}
  \mathbb{1}_{3\times 3} & - \pmb{ \epsilon}_\chi \\ \hline
 \pmb{ \epsilon}^\dagger_\chi  &  \mathbb{1}_{2\times 2}
 \end{array} \right)_{5\times 5}\,, \label{eq:VI_II}
  \end{align}
  where we only retain the leading contributions, and  the effects, which are smaller than $\pmb{\epsilon_\chi}$  with $\chi=R,L$, have been dropped, such as $\pmb{\epsilon^\dagger_\chi \epsilon_\chi}$,   $\pmb{m}_{\ell} \pmb{\delta m}_{1,2}/\pmb{m}^2_{L}$, etc.  The explicit expressions of $\pmb{\epsilon_\chi}$   are  given as:
 \begin{align}
\pmb{ \epsilon^\dagger_L} & = \left(\begin{array}{ccc}
 0 \,, & \frac{v_S y'_\mu }{\sqrt{2} m_{4L} }\,, &  0 \\
  0\,, &   0 \,, &  \frac{v_S y'_\tau }{\sqrt{2} m_{5L} }
 \end{array} \right)  \,, ~~ 
 \pmb{\epsilon^\dagger_R } = \left(\begin{array}{ccc}
 0 \,, & 0\,, &  \frac{v_H y_\tau}{\sqrt{2} m_{4L}}  \\
  0\,, &   \frac{v_H y'_\mu }{\sqrt{2} m_{5L}} \,, & 0
 \end{array} \right)  \,,
 \end{align}
 where the Yukawa couplings $y_{\mu, \tau}$ and $y'_{\mu, \tau}$ are taken as  real numbers.  If we use $v_S \approx 100$ GeV, $m_{4L}\approx m_{5L} \approx 1000$ GeV, and $y^{(\prime)}_{\mu(\tau)} \sim 0.1$, the off-diagonal mixing matrix elements of $U^\chi_\ell$ are of $O(10^{-2})$.  Comparing with the PMNS matrix, where the minimal element is ${\bf U}_{13}\sim 0.14$ and is one oder of magnitude larger than $(U^\chi_\ell)_{ij}$ with $i\neq j$, we can approximate the PMNS matrix to be ${\bf U} \equiv  U^L_\nu U^{L\dagger}_\ell \approx U^L_\nu$. That is, in this leading order approximation, we can use the PMNS matrix to diagonalize the induced neutrino mass matrix. 
 
  After rotating the lepton weak states to physical states based on the $U^{R}_\ell$ and $U^L_\ell$, the Yukawa couplings of the SM Higgs  to the charged leptons  are expressed as:
 \begin{align}
 -{\cal L}_{h} &= \left(\begin{array}{cc}
    \pmb{\bar \ell_L }\,, & \pmb{\bar \Psi}_{\ell L} \end{array} \right) U^L_\ell \left( \begin{array}{c|c}
  ~~\pmb{ m_{\ell} }_{3\times 3}~~ & ~~0~~ \\ \hline
 \pmb{ \delta m^T_2 }  & ~~0~~
 \end{array} \right) U^{R ^\dagger}_\ell
  \left(\begin{array}{c}
   \pmb{\ell_R} \\
      \pmb{\Psi}_{\ell R} \end{array}\right) \frac{h}{v_H} \,, \label{eq:Lh}
 \end{align}
 where we still use $\pmb{\ell}$ and $\pmb{\Psi}^T_{\ell}$ to represent the charged leptons. As a result, the SM Higgs Yukawa couplings to the light charged leptons can be found as:
  \begin{equation}
  - {\cal L}_{h} \supset  \frac{m_\ell}{v_H} \bar \ell_L \ell_R h - \frac{v_S y'_\mu y_\tau}{2 m_{4L}} \bar \mu_L \tau_R h - \frac{v_S y'_\tau y_\mu}{2m_{5L}} \bar\tau_L \mu_R h + H.c. \label{eq:Ymutauh}
   \end{equation}
 The second and third terms can lead to the $h\to \mu \tau$ decay.

\section{ Majorana neutrino mass matrix and FGM patterns}

In this section, we discuss  the neutrino mass matrix and some phenomenology in our model. 
%
When we write the symmetric Majorana neutrino mass matrix as:
 \begin{equation}
 M^\nu = 
 \left(
\begin{array}{ccc}
 m_{ee}  &  m_{e\mu} & m_{e \tau}   \\
 m_{e\mu}  &  m_{\mu\mu} &  m_{\mu \tau} \\
 m_{e\tau} & m_{\mu \tau}  & m_{\tau\tau}   
\end{array}
\right)\,, \label{eq:Mnu}
 \end{equation}
 from the Yukawa couplings in Eqs.~(\ref{eq:Yukawa1}) and (\ref{eq:lang}), each matrix element can then be expressed as:
  \begin{align}
  m_{ee} & = \frac{Y_{ee} v_\Delta}{\sqrt{2}}+ \frac{Y'_{45} y^*_e y'^*_e v^2_{S'} v_{\Delta} }{2\sqrt{2} m_{4L} m_{5L}}\,, \ m_{e\mu}= \frac{y^*_e Y_{\mu 4} v_{S'} v_\Delta }{\sqrt{2} m_{4L}} + \frac{Y'_{45} y'^*_e  y'^*_\mu}{ 2  \sqrt{2} } \frac{v_S v_{S'} v_\Delta}{ m_{4L} m_{5L} }   + \frac{Y'_{45} y^*_e m_{5\mu}}{2} \frac{v_{S'} v_\Delta}{m_{4L} m_{5L}}  \,,  \nonumber \\
 m_{e\tau} & = \frac{y'_e Y_{\tau 5} v_{S'} v_\Delta }{\sqrt{2} m_{5L}} + \frac{Y'_{45}  y^*_e y'^*_\tau}{ 2  \sqrt{2} } \frac{v_S v_{S'} v_\Delta}{ m_{4L} m_{5L} }  + \frac{Y'_{45} y'^*_e m_{4\tau}}{2} \frac{v_{S'} v_\Delta}{m_{4L} m_{5L}}  \,, \  m_{\mu\mu}   = \frac{Y_{\mu 4} y'^*_\mu   v_{S} v_{\Delta}}{2 m_{4L}} \,,  \nonumber \\
  m_{\mu \tau}  &= \frac{Y_{\mu\tau} v_\Delta }{\sqrt{2}} + \frac{Y_{\mu 4} m_{4\tau} v_\Delta}{\sqrt{2} m_{4L}} + \frac{Y_{\tau 5} m_{5\mu} v_\Delta}{\sqrt{2} m_{5L}} + \frac{\eta }{2\sqrt{2}} \frac{ v^2_{S} v_{\Delta} }{ m_{4L} m_{5L}}\,, \ m_{\tau\tau}  = \frac{Y_{\tau 5} y'^*_\tau v_S v_\Delta}{2 m_{5L} }   \label{eq:textures}
  \end{align}
  with $\eta=Y_{\mu 4} y'_S y'^*_\tau + Y_{\tau 5} y_S y'^*_\mu +(Y_{45}+Y'_{45}) y'^*_\tau y'^*_\mu$.
Although the  neutrino mass matrix comes from the dimension-4, -5, and -6 operators, since the involved free parameters are different,  the matrix entries in Eq.~(\ref{eq:textures}) can be taken as the same  order of magnitude with no particular hierarchy, unless there is a further indication.  Due to the $U(1)_{L_\mu -L_\tau}$ gauge symmetry, the light charged-lepton mass matrix in the first term of  Eq.~(\ref{eq:Yukawa2})  is diagonal.  Although the other Yukawa interactions  can induce  off-diagonal elements, as shown earlier, these induced terms indeed are suppressed. If  we neglect these small  off-diagonal effects as a leading approximation,  the Majorana neutrino mass matrix can be diagonalized by the PMNS matrix as $M^\nu_{\rm dia} = {\rm diag}( m_1,  m_2,  m_3)={\rm diag}(|m_1| e^{- i\alpha_{13}},  |m_2| e^{- i\alpha_{23}},  |m_3|   ) = {\bf U}^T M^\nu {\bf U}$, where $\alpha_{13}$ and $\alpha_{23}$ are the  Majorana CP violating phases, 
 and the standard parametrization of PMNS matrix is given as~\cite{PDG}: 
\begin{equation}
{\bf U} =  \begin{pmatrix} 
c_{12} c_{13} & s_{12} c_{13} & s_{13} e^{-i \delta} \\
-s_{12} c_{23} - c_{12} s_{23} s_{13} e^{i \delta} & c_{12} c_{23} -s_{12} s_{23} s_{13} e^{i \delta} & s_{23} c_{13} \\
s_{12} s_{23} - c_{12} c_{23} s_{13} e^{i \delta} & -c_{12} s_{23} - s_{12} c_{23} s_{13} e^{i \delta} & c_{23} c_{13} 
\end{pmatrix}  \label{eq:PMNS}
\end{equation}
with $s_{ij} \equiv \sin \theta_{ij}$, $c_{ij} \equiv \cos \theta_{ij}$, and $\delta$ being the Dirac CP violating phase. 

 From Eq.~(\ref{eq:Mnu}),  there are six different complex matrix elements. After rotating three unphysical phases, we have nine independent parameters.  Since neutrino oscillation experiments cannot observe the two Majorana CP phases, even  we assume $\alpha_{13}=\alpha_{23}=0$,  there are  seven free parameters. However,  we only have six observables: $\Delta m^2_{21, 31}$, $\sin^2\theta_{12,13,23}$, and Dirac CP phase $\delta$; that is, we cannot determine all free parameters without further theoretical or experimental inputs.  It has been suggested that  a class of neutrino mass matrices may suffice to explain all neutrino experiments  if the matrix textures  have two independent zeroes~\cite{Frampton:2002yf}. The seven possible Frampton-Glashow-Marfatia (FGM) matrix patterns are classified  as:
 \begin{align}
& \mathbf{A}_1:
\begin{pmatrix} 0 & 0 & X \\ 0 & X & X \\
X & X & X
\end{pmatrix}
\,,\quad
\mathbf{A}_2:
\begin{pmatrix}
0 & X & 0 \\ X & X & X \\ 0 & X & X
\end{pmatrix}
\,,\quad
\mathbf{B}_1:
\begin{pmatrix}
X & X & 0 \\ X & 0 & X \\ 0 & X & X
\end{pmatrix}\,, \quad \mathbf{B}_2:
\begin{pmatrix}
X & 0 &X \\ 0 & X & X \\ X & X & 0
\end{pmatrix}\,, \nonumber \\
& \mathbf{B}_3:
\begin{pmatrix}
X & 0 & X \\ 0 & 0 & X \\ X & X & X
\end{pmatrix}\,,\quad
\mathbf{B}_4:
\begin{pmatrix}
X & X & 0 \\ X & X & X \\ 0 & X & 0
\end{pmatrix}\,, \quad \mathbf{C}:
\begin{pmatrix}
X & X & X \\ X & 0 & X \\ X & X & 0
\end{pmatrix}\,, \label{eq:FGM}
\end{align}
where the symbol $X$ denotes a nonzero texture. A detailed study with two-zero textures can be found in~\cite{Fritzsch:2011qv,Meloni:2012sx,Ludl:2014axa}. In order to simplify the  analysis, we thus  employ the FGM patterns as the theoretical inputs.  

As mentioned earlier, the neutrino mass order is still uncertain, i.e. $|m_1|< |m_2|<|m_3|$ or $|m_{3}| < |m_1| < |m_2|$ is allowed. With an FGM pattern, it helps understand what form of a neutrino mass matrix  can lead to a specific mass order.   According to  the study referenced in~\cite{Cebola:2015dwa},  by taking the neutrino  data with $1\sigma$ errors,  the NO spectrum could be achieved by the patterns ${\bf A}_{1,2}$ and ${\bf B}_{1,2,3,4}$, while the IO could be achieved by  the patterns ${\bf B}_{1,3}$ and ${\bf C}$. Accordingly, it is of interest to see how the  matrix elements of  Eq.~(\ref{eq:textures}) in our model  connect to those of a specific FGM matrix.  It is found that when some Yukawa couplings are required to vanish, a definite FGM matrix pattern can then be achieved. We show the  vanishing Yukawa couplings for the corresponding FGM matrix  in Table~\ref{tab:neutrino}.  
 It is worth mentioning that  a powerful FGM matrix pattern can also predict the absolute values of neutrino masses  and Majorana CP-phases, which so far have not yet been observed in experiments.
 From the zero textures $M^\nu_{ij}=M^\nu_{k l}=0$ ($ij \neq k l$), the neutrino mass ratios and Majorana CP phases can be  obtained as~\cite{Fritzsch:2011qv}:
 \begin{align}
 \frac{|m_1|}{|m_3|} & = \left| \frac{U_{i3} U_{j3} U_{k 2} U_{l 2} - U_{i2} U_{j2} U_{k 3} U_{l 3}}{U_{i2} U_{j2} U_{k 1} U_{l 1} - U_{i1} U_{j1} U_{k 2} U_{l 2}} \right|\,, \nonumber \\
  \frac{|m_2|}{|m_3|} &= \left| \frac{U_{i1} U_{j1} U_{k 3} U_{l 3} - U_{i 3} U_{j 3} U_{k  1} U_{l 1}}{U_{i 2} U_{j 2} U_{k 1} U_{l 1} - U_{i 1} U_{j 1} U_{k  2} U_{l 2}} \right| \nonumber \\
  \alpha_{13} & =  {\rm arg} \left[\frac{U_{i3} U_{j3} U_{k 2} U_{l 2} - U_{i2} U_{j2} U_{k 3} U_{l 3}}{U_{i2} U_{j2} U_{k 1} U_{l 1} - U_{i1} U_{j1} U_{k 2} U_{l 2}} \right]\,, \nonumber \\
\alpha_{23} & =  {\rm arg} \left[  \frac{U_{i1} U_{j1} U_{k 3} U_{l 3} - U_{i 3} U_{j 3} U_{k  1} U_{l 1}}{U_{i 2} U_{j 2} U_{k 1} U_{l 1} - U_{i 1} U_{j 1} U_{k  2} U_{l 2}} \right]\,.
\end{align} 
 The values of  the neutrino mass ratios and CP phases for each pattern with two benchmark inputs are shown in Table~\ref{tab:predictions}, where  in addition to the taken values of $\sin^2 \theta_{12} =0.304$ and $\sin^2 \theta_{13} =0.0219$, the values inside brackets correspond to  two different inputs: for the left value, we fix $\delta = 1.5 \pi$ and  $\sin^2 \theta_{23} = 0.5$; for the right, $\delta = 1.59205 \pi$ and $\sin^2 \theta_{23} = 0.4515$ are used. From the results, it can be seen that the patterns ${\bf A_1}$ and ${\bf A_2}$ prefer the  normal hierarchy, and the patter ${\bf C}$ shows the inverted hierarchy and degenerate case. The mass ordering   in patterns ${\bf B_{1-4}}$ depends on the taken parameters.  For illustration, in the following analysis, we focus the detailed analysis on the patterns ${\bf A}_1$ and ${\bf C}$.

\begin{table}[tb]
\caption{  Vanishing Yukawa (VY) couplings to determine the FGM two-zero textures in the model.}
\begin{tabular}{c|cccc} \hline \hline
    Pattern    &  ~~~${\bf A}_1$~~~ & ~~~${\bf A}_2$~~~ & ~~~${\bf B}_1$~~~ & ~~~${\bf B}_2$~~~   \\ \hline 
 VY  & ~~$(Y_{ee},~Y'_{45},~ y_e) \approx 0$~~  & ~~$(Y_{ee},~Y'_{45},~y'_e)\approx 0$~~ & ~~$(y'_e,~Y'_{45},~y'_\mu)\approx 0$~~ & ~~$(y_e,~Y'_{45},~Y_{\tau 5})\approx 0$~~  \\ \hline
  Pattern & ~~~${\bf B}_3$~~~ & ~~~${\bf B}_4$~~~ & ${\bf C}$    \\ \hline 
  VY & ~~~$(y'_e, ~y_e,~Y_{\mu 4}) \approx 0$ ~~~& ~~~$(y_e,~Y_{\tau 5},~m_{4\tau}) \approx 0$~~~ & ~~~$(Y_{\mu4},~Y_{\tau 5})\approx 0$~~~  \\ \hline
\end{tabular}
\label{tab:neutrino}
\end{table}

\begin{table}[tb]
\caption{  Mass ratios and Majorana CP phases of each FGM  pattern with some benchmark inputs, where in addition to the taken values of $\sin^2 \theta_{12} =0.304$ and $\sin^2 \theta_{13} =0.0219$, the values inside brackets correspond to two different inputs: for the left value, we fix $\delta = 1.5 \pi$ and  $\sin^2 \theta_{23} = 0.5$; for the right, $\delta = 1.59205 \pi$ and $\sin^2 \theta_{23} = 0.4515$ are taken. }
\begin{tabular}{c|cc} \hline \hline
 & ~~~mass relation~~~ & ~~~CP-violating phases~~~ \\ \hline
~${\bf A}_1$~ & ~$\frac{|m_1|}{|m_3|} \simeq (0.10, \, 0.087) $, \ $\frac{|m_2|}{|m_3|} \simeq (0.23, \, 0.22)$~ & $\alpha_{13} \simeq (0.43 \pi, \, 0.33 \pi), \ \alpha_{23} \simeq (-0.47 \pi, \, -0.56 \pi)$  \\ \hline
~${\bf A}_2$~ & ~$\frac{|m_1|}{|m_3|} \simeq (0.10, \, 0.12) $, \ $\frac{|m_2|}{|m_3|} \simeq (0.23, \, 0.25) $~ & $\alpha_{13} \simeq (-0.43 \pi, \, -0.53 \pi), \ \alpha_{23} \simeq (0.47 \pi, \, 0.38 \pi)$  \\ \hline
~${\bf B}_1$~ & ~$\frac{|m_1|}{|m_3|} \simeq (1.0, \, 0.95) $, \ $\frac{|m_2|}{|m_3|} \simeq (1.0, \, 0.74)$~ & $\alpha_{13} \simeq (1.0 \pi, -0.98 \pi), \ \alpha_{23} \simeq (-1.0 \pi, \, -0.99 \pi)$  \\ \hline
~${\bf B}_2$~ & ~$\frac{|m_1|}{|m_3|} \simeq (1.0, \, 1.1)$, \ $\frac{|m_2|}{|m_3|} \simeq (1.0, \, 1.3)$~ & $\alpha_{13} \simeq (-1.0 \pi, \, -0.98 \pi), \ \alpha_{23} \simeq (1.0 \pi, \, -0.99 \pi)$  \\ \hline
~${\bf B}_3$~ & ~$\frac{|m_1|}{|m_3|} \simeq (1.0, \, 0.73)$, \ $\frac{|m_2|}{|m_3|} \simeq (1.0, \, 0.87)$~ & $\alpha_{13} \simeq (-1.0 \pi, \, 0.98 \pi), \ \alpha_{23} \simeq (-1.0 \pi, \, -1.0 \pi)$  \\ \hline
~${\bf B}_4$~ & ~$\frac{|m_1|}{|m_3|} \simeq (1.0, \, 1.4 )$, \ $\frac{|m_2|}{|m_3|} \simeq (1.0, \, 1.1)$~ & $\alpha_{13} \simeq (1.0 \pi, \, 0.98 \pi), \ \alpha_{23} \simeq (1.0 \pi, \, -1.0 \pi)$  \\ \hline
~${\bf C}$~ & ~$\frac{|m_1|}{|m_3|} \simeq (1.0, \,1.19)$, \ $\frac{|m_2|}{|m_3|} \simeq (1.0, \, 1.2)$~ & $\alpha_{13} \simeq (1.0 \pi, \, 0.70 \pi), \ \alpha_{23} \simeq (1.0 \pi, \, -0.89 \pi)$  \\ \hline
\end{tabular}
\label{tab:predictions}
\end{table}

\section{Numerical analysis and other phenomena of interest} 

\subsection{ Explain neutrino data  and predict the absolute neutrino masses}

Since our purpose is not to examine  all  FGM patterns, in the following numerical analysis,  we take   ${\bf A}_1$ and ${\bf C}$ as the representatives  of the NO and IO mass spectra, respectively. To determine the non-vanishing entries of the neutrino mass matrix and $|m_i|$, we scan the parameters with  the neutrino  data at the $1\sigma$ level.  Due to large experimental uncertainty, the Dirac CP phase is taken from a global data analysis using an $\chi^2$ method~\cite{Capozzi:2017ipn}, in which the result in the $1\sigma$ region is $\delta/\pi=(1.18, 1.61)$ for NO and $\delta/\pi=(1.12,1.62)$ for IO.  
Combining the experimental inputs with two independent zero textures, we basically  have eight known inputs; thus,  we can completely constrain the four non-vanishing complex entries of  ${\bf A}_1$ and ${\bf C}$.

 Using the relation $M^\nu = {\bf U}^* M^\nu_{\rm dia} {\bf U}^\dagger$ and the zero textures in  $M^\nu$, 
 the mass relations in  ${\bf A}_1$ can be expressed as:
  \begin{align}
  m^*_1 & = \frac{U_{13}}{U_{11}} \left(\frac{ U_{12} U_{23} -U_{13} U_{22} } {U_{11} U_{22} -U_{12} U_{21}} \right) m^*_3 \,, \nonumber \\
  m^*_2 & = -\frac{U_{13}}{U_{12}} \left(\frac{ U_{11} U_{23} -U_{13} U_{21} } {U_{11} U_{22} -U_{12} U_{21}} \right) m^*_3, \label{eq:mA}
  \end{align}
 while in  {\bf C},  they are:
  \begin{align}
    m^*_1 & =  \frac{ U^2_{22} U^2_{33} - U^2_{23} U^2_{32} }{  U^2_{21}  U^2_{32} - U^2_{22} U^2_{31}  } m^*_3\,, 
  \nonumber \\
 m^*_2 & =  - \frac{U^2_{21} U^2_{33} - U^2_{23} U^2_{31}  }{ U^2_{21}  U^2_{32} - U^2_{22} U^2_{31} } m^*_3\,,  \label{eq:mC}
  \end{align}
 where the $m_k$s values in general are complex; however, there are only two independent phases among $m_{1,2,3}$.
 With the chosen Majorana  phases, such as $m_1 = |m_1| e^{-i\alpha_{13}}$ and $m_2 = |m_2| e^{- i \alpha_{23}}$, we obtain the relations
 \begin{align}
 \alpha_{13} =  {\rm arg}\left[ \frac{U_{13}}{U_{11}} \left(\frac{ U_{12} U_{23} -U_{13} U_{22} } {U_{11} U_{22} -U_{12} U_{21}} \right)\right]\,, \quad 
\alpha_{23} =  {\rm arg} \left[ -\frac{U_{13}}{U_{12}} \left(\frac{ U_{11} U_{23} -U_{13} U_{21} } {U_{11} U_{22} -U_{12} U_{21}} \right)\right] 
 \end{align}
 for the ${\bf A}_1$ case, and 
 \begin{align}
 \alpha_{13} = {\rm arg}\left[ \frac{ U^2_{22} U^2_{33} - U^2_{23} U^2_{32} }{  U^2_{21}  U^2_{32} - U^2_{22} U^2_{31}  } \right], \quad 
 \alpha_{23} = {\rm arg} \left[ - \frac{U^2_{21} U^2_{33} - U^2_{23} U^2_{31}  }{ U^2_{21}  U^2_{32} - U^2_{22} U^2_{31} } \right] 
 \end{align}\
 for the ${\bf C}$ case.
 If we take the central values of measured $\theta_{12, 13}$  in Eq.~(\ref{eq:nu_data}), $\sin^2\theta_{23} \approx 0.50$, and $\delta \approx 1.5 \pi$,  we can easily obtain: 
  \begin{equation}
  {\bf A}_1 : \left\{
\begin{array}{c}
  |m_1|/|m_3| \approx  0.230 \,, \\
    |m_2|/|m_3| \approx  0.102 \,, \\
|m_2|^2 - |m_1|^2 \approx 0.029 |m_3|^2 \,, \\
\alpha_{13} \approx 0.430 \pi \,, \\
\alpha_{23} \approx -0.469 \pi \,.
\end{array}
\right.  \label{eq:A1}
  \end{equation}
  However, it is found that the pattern ${\bf C}$ is very sensitive to the values of the mixing angles and CP phase $\delta$ when $\Delta m^2_{21}$ and $\Delta m^2_{32}$ are required to fit the data within $1\sigma$ errors. If $\sin^2\theta_{23}\approx 0.4515$ and $\delta \approx 1.59205\pi$ are taken,  we obtain:
\begin{equation}
{\bf C} : \left\{
\begin{array}{c}
   |m_1|/|m_3| \approx  1.19 \,, \\
     |m_2|/|m_3| \approx  1.20 \,, \\
|m_2|^2 - |m_1|^2 \approx 0.0130 |m_3|^2 \,, \\
\alpha_{13} \approx -0.705 \pi \,, \\
\alpha_{23} \approx 0.887 \pi \,.
\end{array}
\right.  \label{eq:C}
  \end{equation} 
Accordingly, if we further take  $\Delta m^2_{21} \approx 7.53 \times 10^{-5}$ eV$^2$, the values of $|m_i|$ and $\Delta m^2_{23}$ can be determined  as:
  \begin{equation}
 { \bf A}_1 : \left\{
\begin{array}{c}
  |m_1|  \approx 5.5\times 10^{-3} \ {\rm eV}\,, \\
  |m_2|  \approx 1.03 \times 10^{-2} \ {\rm eV}\,, \\
  |m_3|   \approx  5.06 \times 10^{-2}\ {\rm eV} \,, \\
    \Delta m^2_{32}  \approx  2.45\times 10^{-3} \ {\rm eV}^2 \,; \\
\end{array}
\right. \quad
{\bf C} : \left\{
\begin{array}{c}
  |m_3| \approx  7.60 \times 10^{-2} \ {\rm eV} \,, \\
   |m_1| \approx  9.07 \times 10^{-2} \ {\rm eV}\,, \\
    |m_2| \approx  9.11\times 10^{-2} \ {\rm eV}  \,, \\
     \Delta m^2_{23} \approx  2.53\times 10^{-3} \ {\rm eV}^2 \,.
     \end{array}
\right.  \label{eq:pred}
  \end{equation}

 From above analysis,   ${\bf A}_1$ and ${\bf C}$ can fit the neutrino  data for the NO and IO mass spectra at the $1\sigma$ level, respectively. However, if we compare the results with the cosmological   limit on the sum of neutrino masses, which is given as:
 \begin{equation}
 \sum_\nu m_\nu < (0.12, 0.17) \ {\rm eV}\,, \ \text{(\cite{Vagnozzi:2017ovm}, \cite{Couchot:2017pvz})}
 \end{equation}
 it can be found that the resulting $\sum_j |m_{j}|$ in  ${\bf A}_1$ can satisfy the upper bound while that in  ${\bf  C}$ is  higher than the limit. 
 In order to determine whether the tension with the  cosmological neutrino mass bound can be relaxed when the ranges of the experimental measurements are extended,  we adopt neutrino data  up to the $3\sigma$ level instead of those at the $1\sigma$ level for ${\bf C}$.  In the numerical analysis, we generate $5 \times 10^8$ sampling points by randomly selecting the experimental values of $s_{12,23,13}$ and $\delta$ within  $\{1\sigma,2\sigma,3\sigma \} $ errors and the values of $m_{1}$ in the range of $[0.01, 0.17]$ eV; then, $m_2$ and $m_3$ are obtained via Eq.~(\ref{eq:mC}).  In the end, the number of output points, which  can fit the $\Delta m_{21,23}^2$ data in the  $\{1\sigma,2\sigma,3\sigma \} $ range, is $\{ 552, 3004, 3467\}$. The obtained Dirac CP phase $\delta$ and $\sum_j |m_j|$ are shown in Fig.~\ref{fig:C_2sigma}, where the dots  in black, red and blue denote the results with $1\sigma$, $2\sigma$ and $3\sigma$ errors, respectively. From the figure, it can be clearly seen that $\sum_j |m_j|$ in  {\bf C} is excluded even at the $3\sigma$ level if we adopt the bound from the cosmological measurements $\sum_\nu m_\nu < 0.12$ eV while it can still satisfy the bound at the $2\sigma$ level if we adopt the upper limit of $0.17$ eV.
\begin{figure}[hptb] 
\begin{center}
\includegraphics[width=80mm]{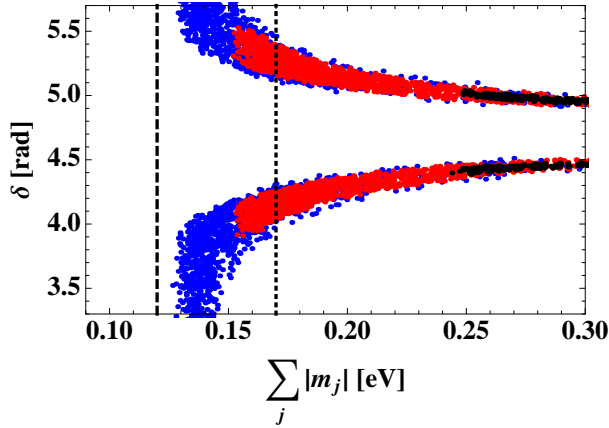} 
\caption{  Scatter plot for the Dirac CP phase and $\sum_j |m_j|$, where the dots in black, red, and blue denote the neutrino data with $1\sigma$, $2\sigma$, and $3\sigma$ errors, respectively. 
 The dashed (dotted) line denote the cosmological neutrino mass bound $0.12(0.17)$ eV. }
\label{fig:C_2sigma}
\end{center}
\end{figure}

Since the uncertainties of $\sin^2\theta_{23}$ and $\Delta m^2_{32}$ in Eq.~(\ref{eq:nu_data}) correspond to a $68\%$ confidence level (CL), and  the pattern ${\bf C}$ cannot  fit the data within $1\sigma$ errors, in the remaining part of the paper,
we only use the pattern ${\bf A}_1$ to show the constraints for the relevant Yukawa couplings. From the mass diagonal relation $M^\nu_{\ell \ell'} = \left( U_{\ell k} U_{\ell' k}\right)^* m_k$,  when the PMNS matrix entries and $m_k$ are known, $M^\nu_{\ell \ell'}$ can then be determined.  Thus,   the correlation between $\delta$ and $|m_j|$ in  ${\bf A}_1$ is shown in Fig.~\ref{fig:Mnu_A1}(a), where the neutrino data within $1\sigma$ error have been included. From the plot, it can be seen that each $|m_i|$ narrowly spreads around  the value of Eq.~(\ref{eq:pred}). In the plot, we also show the effective Majorana neutrino mass $\langle m_{\beta\beta}\rangle$, which is related to the neutrinoless double-beta ($0\nu\beta\beta$) decay rate and is defined by~\cite{Asai:2017ryy}:
 \begin{equation}
 \langle m_{\beta\beta}\rangle = |\sum_k U^2_{ek} m_k |\,,
 \end{equation}
 where a $90\%$ CL upper limit  of $\langle m_{\beta\beta} \rangle < 0.061-0.165$  eV was obtained by the KamLAND-Zen collaboration~\cite{KamLAND-Zen:2016pfg}. Our result of $\langle m_{\beta\beta} \rangle \approx ( 0.34,\, 2.3) \times 10^{-2}$ eV clearly satisfies the bound. According  the results, the allowed ranges of $|m_{ij}|$ as a correlation of $|m_{\tau \tau}|$  are shown in Fig.~\ref{fig:Mnu_A1}(b), where  we scan the parameters using $10^{7}$ sampling points to fit the neutrino data, and  $|m_1| \in [0.001, 0.1]$ eV is taken. As a result, the obtained ranges of $m_{ij}$ in ${\bf A}_1$ are given as: 
 \begin{align}
 m_{e\tau} & =(0.99, 1.11)\times 10^{-2} \, {\rm eV}\,, \quad  m_{\mu \mu}=(2.5,3.0)\times 10^{-2} \, {\rm eV}\,, \nonumber \\ 
 m_{\mu \tau}& =( 2.2, 2.4)\times 10^{-2} \, {\rm eV}\,, \quad  m_{\tau \tau}=(2.4, 2.8)\times 10^{-2} \, {\rm eV}\,,  \label{eq:mij}
 \end{align}
 where $m_{ee}$ and $m_{e \mu}$ are zero in neutrino mass pattern $\mathbf{A}_1$.
  In addition, the correlation between the Dirac phase $\delta$ and Majorana phase $\alpha_{13} [\alpha_{23}]$ are shown in Fig.~\ref{fig:Mnu_A1}(c)[(d)].
\begin{figure}[hptb] 
\begin{center}
\includegraphics[width=75mm]{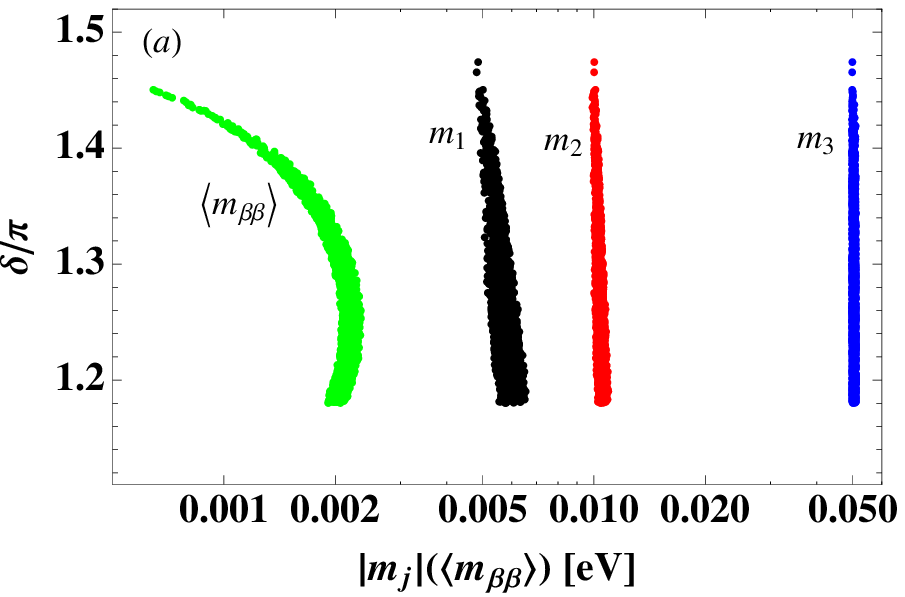} 
\includegraphics[width=80mm]{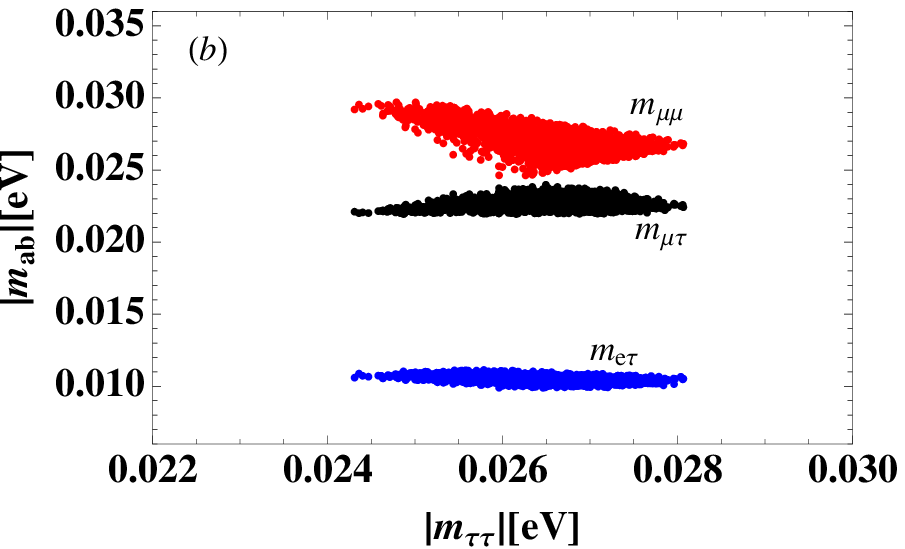} 
\includegraphics[width=75mm]{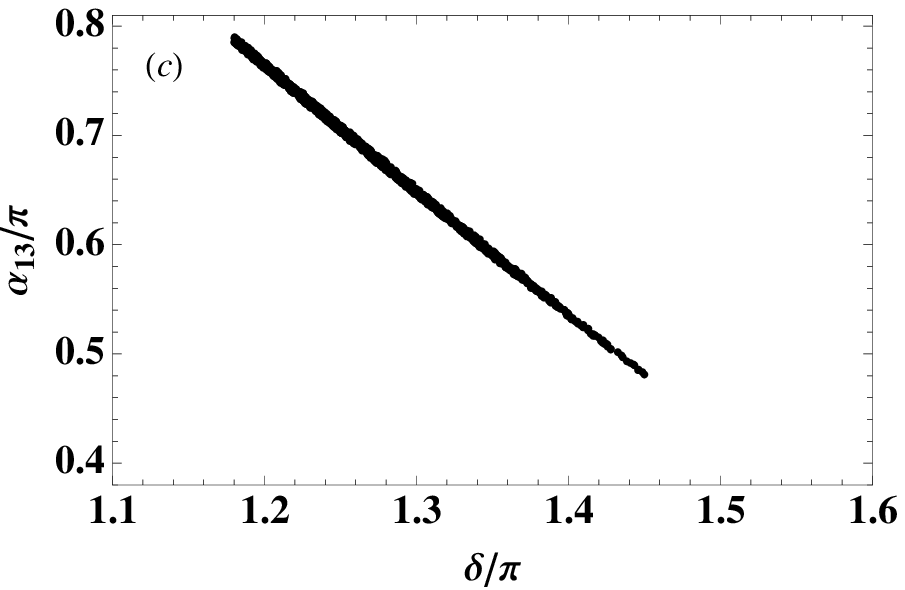} 
\includegraphics[width=75mm]{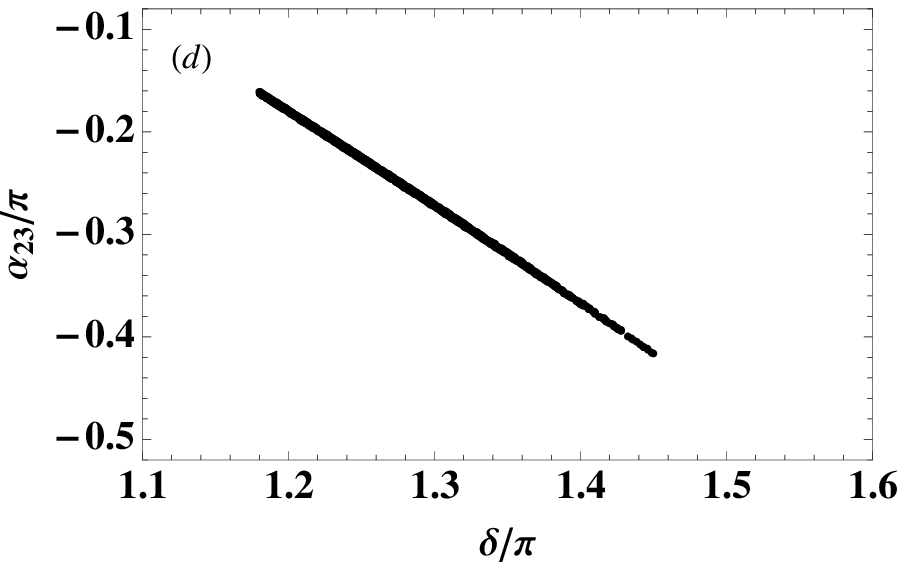} 
\caption{ (a) Predicted $|m_j|$ and effective Majorana neutrino mass for the $0\nu\beta\beta$ decay; 
(b) allowed ranges for $|m_{ij}|$ as a correlation of $|m_{\tau \tau}|$; (c)[(d)] correlation between Dirac phase $\delta$ and Majorana phase $\alpha_{13}[\alpha_{23}]$, where FGM  pattern ${\bf A}_1$ is applied, and neutrino data within $1\sigma$ errors are taken.  }
\label{fig:Mnu_A1}
\end{center}
\end{figure}
 
 \subsection{ Limits of Yukawa couplings and  the $h\to \mu \tau$ decay} 
 
 Based on the  results obtained above, we now discuss the limits on the introduced Yukawa couplings shown in Eqs.~(\ref{eq:Yukawa1}) and (\ref{eq:Yukawa2}).  To simplify the analysis, we take $m_{4L}\approx  m_{5L} \equiv m_{L}$ and $v_{S}\approx v_{S'} \equiv v_{X}$, and define the parameters as:
 \begin{align}
 a_L & = \frac{ y'^*_\tau y^*_\mu v_X }{2 m_L}\,, \quad a_R = \frac{y'_\mu y_\tau v_X}{2m_L}\,, \quad \xi^{(\prime)}_{ab} = \frac{Y^{(\prime)}_{ab} v_\Delta}{\sqrt{2}}\,.
 \end{align}
 The parameters $a_{R,L}$ can lead to the Higgs lepton-flavor violating $h \to \mu \tau$ decay, where  the associated interactions from Eq.~(\ref{eq:Ymutauh}) are expressed as~\cite{Dorsner:2015mja,Herrero-Garcia:2016uab}:
  \begin{equation}
   {\cal L}_{h\tau\mu} =  h \bar \mu (a_R P_R+ a_L P_L)  \tau + H.c. \label{eq:taumuh}
  \end{equation}
The BR for $h\to \tau \mu$ can be obtained as:
  \begin{equation}
  \label{eq:aLaR}
  BR(h\to \mu \tau)= \frac{|a_L|^2 + |a_R|^2}{8 \pi \Gamma_h} m_h\,.
  \end{equation}
  With $m_h\approx 125$ GeV and $\Gamma_h\approx 4.21$ MeV, the limit on $a_{L,R}$ can be obtained as 
  \begin{equation}
  \sqrt{|a_L|^2 +|a_R|^2} \approx 1.56\times 10^{-3} \sqrt{\frac{BR(h\to  \tau \mu)}{2.5\times 10^{-3}}}\,, 
  \end{equation}
 where $BR(h\to  \mu \tau)$ can be taken from the experimental data, and  the current upper limits from ATLAS and CMS are $1.43\%$~\cite{Aad:2016blu} and $0.25\%$ ~\cite{Khachatryan:2015kon,CMS:2017onh}, respectively.

  Using $Y_{ee}\approx Y'_{45}\approx y_e\approx 0$ in ${\bf A}_1$, the neutrino mass matrix entries in Eq.~(\ref{eq:textures}) are formed as:
 \begin{align}
 m_{e\tau} &=  \frac{ y'_e v_X}{m_L} \xi_{\tau5}\,, \quad  m_{\mu \mu} = \frac{\sqrt{2}} {y^*_\tau} a^*_R \xi_{\mu4}\,, \quad m_{\tau \tau} =  \frac{\sqrt{2}}{y_\mu^*} a_L \xi_{\tau 5}\,, \nonumber \\
 m_{\mu \tau} & = \xi_{\mu \tau} +  \left( \frac{m_{4\tau}}{m_L} + \frac{y'_S}{y^*_\mu} \frac{v_X}{m_L} a_L \right) \xi_{\mu 4}  \nonumber \\ 
 & + \left( \frac{m_{5\mu}}{m_L} + \frac{y_S}{y^*_\tau} \frac{v_X}{m_L} a^*_R \right) \xi_{\tau 5} + \frac{2}{y^*_\tau y^*_\mu} a_L a^*_R (\xi_{45}+\xi'_{45})\,.
 \end{align}
 In order to get sizable $BR(h\to \mu \tau)$ and $\xi_{45}$, we find  that $|a_L|\ll |a_R|$ or $|a_R| \ll  |a_L|$ has to be satisfied.  
  According to Eq.~(\ref{eq:mij}),  if we take $|m_{\mu \mu}| \approx |m_{\tau \tau}| \approx 2.7 \times 10^{-2}$ eV,  $|m_{e\tau}|\approx  10^{-2}$ eV, $|m_{\mu\tau}| \approx 2.3 \times 10^{-2}$ eV, $|a_{R(L)}|\approx 10^{-3}(10^{-8})$,    $v_X\approx 100$ GeV, and $m_L\approx 1000$ GeV, we can obtain  $BR(h\to \mu \tau) \approx 1.2\times 10^{-3}$, and the magnitudes of parameters are obtained as:
  \begin{align}
& |y'_e \xi_{\tau5}| \approx  1.0 \times 10^{-10}\, {\rm GeV} \,, \  \left|\frac{\xi_{\mu 4}}{y_\tau}\right| \approx  1.9 \times 10^{-8}\, {\rm GeV}\,, \nonumber \\
& \left|\frac{\xi_{\tau 5}}{y_\mu}\right| \approx  1.9 \times 10^{-3}\, {\rm GeV}\,,  \ |\xi_{\mu \tau}|  \approx \left|2.3 - \frac{2 \xi_{45}}{y^*_\mu y^*_\tau } \right| \times 10^{-11}\, {\rm GeV}\,, \label{eq:xiv}
  \end{align}
  where the second and third terms in $m_{\mu \tau}$ have been ignored due to  $y_S, y'_{S}, m_{4\tau,5\mu}/m_L \ll 1$. With  $y_\mu \approx  y_\tau \approx  0.1$, the Higgs triplet Yukawa couplings  then have the hierarchy $Y_{\mu\tau} \ll Y_{\mu4} \ll Y_{\tau 5} \ll Y_{45} $; that is,  we cannot avoid  fine-tuning  the Yukawa couplings to explain the neutrino data in this model. 

 According to above analysis, we see that the  Yukawa couplings, which are not highly suppressed by the neutrino masses, are only $y_\mu$, $y_\tau$, $y'_\mu$, and $Y_{45}$. We need to examine if they will be further constrained by other rare decays. Since the new physics effects occur in the lepton sector, the strict constraints may come from the lepton-flavor violating processes, such as $\tau \to 3 \mu$, $\tau\to (e, \mu )\gamma$, and $\mu \to e \gamma$. From Eq.~(\ref{eq:taumuh}), it is known that $\tau \to 3 \mu$ can be induced through off-shell $h$ decay into the muon pair. The BR for this three-body decay can be expressed as:
 \begin{align}
 BR(\tau\to 3\mu)=\frac{\tau_\tau m^5_\tau }{3\cdot 2^9 \pi^3 m^4_h} | y_{\mu\mu} a_R|^2 \approx 1.2 \times 10^{-7} |a_R|^2 \,,
 \end{align}
where $y_{\mu\mu}=m_\mu/v_H$ is the Higgs coupling to the muon in the SM, and the small $a_L$ has been ignored. Taking $a_R\approx  10^{-3}$, the $h$-mediated $BR(\tau\to 3\mu)$ is much less than the current upper bound of $2.1 \times 10^{-8}$~\cite{PDG}. To induce the rare $\mu \to e \gamma$ process, the new Yukawa couplings  have to couple to the electron. From Eqs. (\ref{eq:Yukawa1}) and (\ref{eq:Yukawa2}), the relevant couplings are $Y_{ee}$, $y_e$, and $y'_e$, however,  $Y_{ee}\approx y_e \approx 0$ and $y'_e \ll 1$ have been used  to fit the neutrino masses. Thus, the rare $\mu \to e \gamma$ process is suppressed in our model. 

Similarly, $\tau \to (e,\, \mu) \gamma$ are suppressed by most Yukawa couplings with the exception of $y_\tau$ and $y'_\mu$, where the associated Feynman diagram is shown in Fig.~\ref{fig:tau_mu_ga}. It can be seen that in addition to $y_\tau$ and $y'_\mu$, the quartic scalar coupling $\lambda_{10}$ involves in the $\tau\to \mu \gamma$ process. As a result, the interaction of the loop induced  $\tau\to \mu \gamma$ can be written as:
 \begin{align}
 {\cal L}_{\tau \mu \gamma} & =- \frac{e m_\tau}{16\pi^2} C_R \bar \mu \sigma_{\mu \nu} P_R \tau F^{\mu \nu}\,, \ C_R =  \frac{\lambda_{10}v_H a_R }{2 m_\tau m^2_L} I(z_h,z_S)\,,  \\
 I(z_h, z_S) & = \int^1_0 dx_1 \int^{x_1}_{0} dx_2 \frac{x^2_2}{ (z_h -(z_h-z_S) x_1 + (1-z_S)x_2)^2 }\,, \nonumber 
 \end{align}
with $z_h=m^2_h/m^2_L$ and $z_S=m^2_S/m^2_L$. The BR for $\tau \to \mu \gamma$ can be expressed as:
 \begin{equation}
 \frac{BR(\tau\to \mu\gamma)}{BR(\tau \to \mu \bar \nu_\mu \nu_\tau)}  = \frac{3 \alpha_e}{ 4 \pi G_F} |C_R|^2 \approx 1.51\times 10^{-13} \left( \frac{|a_R| \lambda_{10}}{10^{-3}}\right)^2\,,
 \end{equation}
 where we have used  $a_R\approx 10^{-3}$, $m_h\approx 125$ GeV, $m_S \approx 10$ GeV, $m_L\approx1000$ GeV, and $I(z_h, z_S) \approx 0.46$. Clearly, the BR for $\tau \to \mu \gamma$ in our model is still  below the current upper bound of $4.4\times 10^{-8}$~\cite{PDG}.  Note that we have ignored the $a_L$ effect due to the use of  $a_L \ll a_R$.
\begin{figure}[hptb] 
\includegraphics[width=85mm]{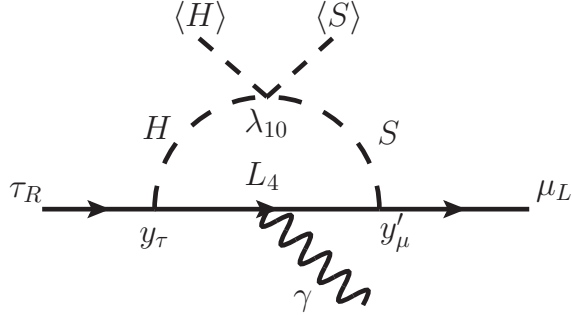}  
\caption{ Sketched Feynman diagram for $\tau\to \mu \gamma$.  }
\label{fig:tau_mu_ga}
\end{figure}

 \subsection{ Phenomenological implications on the muon $g-2$,  rare $\tau$, and $H^{--} \to \mu \tau $ decays}

After determining the magnitudes of the Higgs-triplet Yukawa couplings,  which are used to explain the neutrino data, we state  some  implications of this model in flavor and collider physics,  which have been studied in the literature and are still interesting  in this model.  In addition to the large $BR(h\to \mu\tau)$,  if the new $Z'$ gauge boson is in the MeV to GeV range,  the muon $g-2$ anomaly can be resolved by the  intermediate  $Z'$-gauge boson~\cite{Altmannshofer:2016brv,Altmannshofer:2014cfa,Araki:2017wyg}, depending on the magnitude of $g_{Z'}$ gauge coupling.
 The contribution from the $Z'$-penguin diagram to the muon $g-2$ can be expressed  as~\cite{Altmannshofer:2016brv,Araki:2017wyg}
\begin{align}
\Delta a_\mu^{Z'} = \frac{g_{Z'}^2}{8 \pi^2} \int^1_0 dx \frac{2 m_\mu^2 x^2 (1-x)}{x^2 m_\mu^2 + (1-x)m_{Z'}^2}\,.
\label{eq:delta-mu-g-2}
\end{align} 
It is found that to explain the muon $g-2$ anomaly, $\Delta a_\mu =a^{\rm exp}_\mu - a^{\rm SM}_\mu= (28.7 \pm 8.0)\times 10^{-10}$~\cite{PDG}, the allowed ranges of $g_{Z'}$ and $m_{Z'}$ are :
\begin{align}
2 \times 10^{-4} \leq &~ g_{Z'}  \leq 2 \times 10^{-3}, 
\label{eq:gp-range} \\
5 \leq &~ m_{Z'} \leq 210~ \mathrm{MeV},
\label{eq:mZp-range}
\end{align}
where other regions have been experimentally  excluded, such as the neutrino trident production~\cite{Altmannshofer:2014pba}, BABAR collaboration~\cite{TheBABAR:2016rlg}, and Borexino experiment~\cite{Bellini:2011rx}.

 With the value of $a_R\sim 10^{-3}$,  the sizable Yukawa couplings $y^{\prime}_\mu$ and $y_\tau$ can induce the lepton-flavor violating interaction $\tau$-$\mu$-$S$ through the mixing between vector-like lepton and $\tau(\mu)$-lepton. From the $S$-$Z'$-$Z'$ interaction, the $\tau \to \mu Z'Z'$ decay can be generated by the mediation of light scalar $S$, and its partial decay rate as a function of $Z'$-pair invariant can be derived as~\cite{Chen:2017cic}:
\begin{align}
 \frac{dBR(\tau \to \mu Z'  Z' )}{dq^2} &  \approx  \frac{m_\tau}{64\pi^2 m_h} \frac{\Gamma_h}{ \Gamma_\tau } BR(h\to \mu \tau) \nonumber \\
  & \times  \frac{ (q^2 -2 m_{Z'}^2)^2 + 8 m^4_{Z'}}{v^4_S m^2_S}\left( 1- \frac{q^2}{m^2_{\tau}} \right)^2 \sqrt{1- \frac{4 m^2_{Z'}}{q^2}}\,.
 \end{align}
 It can be  seen that $\tau \to \mu Z' Z'$ and $h\to \mu \tau$ can be correlated in the model when $m_{Z'}$ is below GeV. 
 We show the $BR(\tau\to \mu Z'Z')$ ( in units of $10^{-9}$) as a function of $BR(h\to \mu \tau)$ ( in units of $10^{-3}$) and $v_S$ in Fig.~\ref{fig:taumuZZ}, where $m_S=10$ GeV and $m_{Z'}=0.2$ GeV are fixed. With 50 ab$^{-1}$ of data accumulated at the Belle II, the sample of $\tau$ pairs can be increased up to around $5\times 10^{10}$, where the sensitivity necessary to observe the LFV $\tau$ decays can reach $10^{-10}-10^{-9}$~\cite{Aushev:2010bq}. Therefore, the $BR(\tau\to \mu Z' Z')$ of $10^{-9}$ allowed in the model could be tested at the Belle II. 
 
\begin{figure}[hptb] 
\includegraphics[width=85mm]{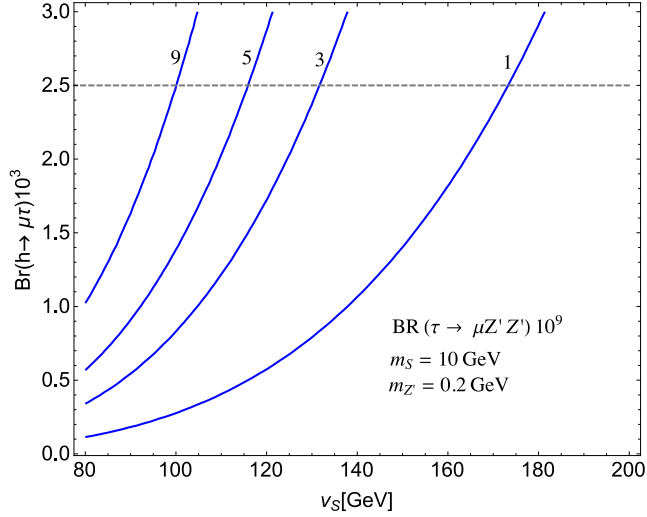}  
\caption{ Correlation between $BR(h\to \mu \tau)$ (in units of $10^{-3})$ and $BR(\tau \to \mu Z'Z')$ (in units of $10^{-9}$, where the horizontal dashed line is the upper bound of $h\to \mu \tau$, and  we have fixed $m_S=10$ GeV and $m_{Z'}=0.2$ GeV.  }
\label{fig:taumuZZ}
\end{figure}

Moreover, we find that a sizable $Y_{45}$ Yukawa coupling can change the decay property of doubly charged Higgs $H^{\pm \pm}$ in the Higgs triplet. In this model, 
$H^{\pm\pm}$ can decay to the  $\mu^\pm \tau^\pm$ final states  via the dimension-4 and the induced dimension-6 operators, which are  expressed as:
 \begin{equation}
Y_{\mu \tau} L^T_\mu C i\tau_2 \Delta L_\tau +  \frac{Y_{45} y_\tau y_\mu}{m^2_L } \tau^T_R H^T i\tau_2 \Delta H \mu_R, 
\label{eq:coupling}
 \end{equation}
 where  the corresponding $H^{\pm\pm}$ Yukawa coupling to $\mu^\pm \tau^\pm$ can be written as:
 \begin{equation}
 Y_{H^{\pm\pm}}= Y^*_{\mu\tau} + Y_{45} y_\tau y_\mu \frac{v^2_H}{2 m^2_L}. 
  \end{equation}
 From Eq.~(\ref{eq:xiv}),  $Y_{45}$ can in principle be $O(0.1)$ when other neutrino mass related parameters are tuned to be small, e.g. $Y_{\mu \tau} v_\Delta /\sqrt{2}\sim 10^{-10}$. Thus, with $m_L \approx 1000$ GeV, $v_H \approx  246$ GeV, $m_{H^{\pm\pm}} \approx 800$ GeV, and $y_\tau \sim y_\mu \sim 0.1$, the decay rate ratio of $H^{\pm\pm} \to \mu^\pm \tau^\pm$ to $H^{\pm\pm} \to W^\pm W^\pm$ can be estimated as~\cite{Han:2007bk}:
  \begin{equation}
  \frac{\Gamma(H^{\pm \pm } \to \mu^\pm \tau^\pm)}{\Gamma(H^{\pm\pm} \to W^\pm W^\pm)} \approx  \frac{|Y_{H^{\pm\pm} }|^2 v^2_H}{2 v^2_\Delta} \frac{v^2_H}{m^2_{H^{\pm\pm}}} \approx  \frac{2.6\times 10^{-4} |Y_{45}|^2 }{v^2_\Delta}\,,
  \end{equation}
  where the small $Y_{\mu\tau}$ is neglected. 
With $|Y_{45}|\sim 0.05$ and $v_\Delta \sim 0.01$ GeV, the ratio can be at the $10\%$ level; that is, the BR for $H^{\pm\pm} \to \mu^\pm \tau^\pm$ is not suppressed and can be a good channel to observe the doubly charged-Higgs. In addition, the $\tau \to \ell_i \ell_j \bar \ell_k$ decays can be induced by the $H^{\pm\pm}$ couplings shown in Eq.~(\ref{eq:coupling}). 
Since we focus on the $\mathcal{A}_1$ pattern, the potential mode is $\tau \to  \mu \mu \bar \mu$ and its BR can be estimated as~\cite{Akeroyd:2009nu}:
\begin{equation}
\frac{BR(\tau \to  \mu \mu \bar \mu)}{BR(\tau \to \mu \bar \nu \nu)} = \frac{1}{4 G_F^2 m^4_{H^{\pm \pm}}} |Y_{H^{\pm \pm}}|^2 \left( \frac{\sqrt{2} m_{\mu \mu}}{v_\Delta} \right)^2. 
\end{equation}
This BR is tiny since it is suppressed by $(m_{\mu \mu}/v_\Delta)^2 \sim 4 \times 10^{-17}$ when  $v_\Delta \sim 0.01 $ GeV is used; therefore, this process cannot give a strict constraint on $Y_{H^{\pm \pm}}$.

\section{Summary}

We studied the origin of the neutrino mass in the gauged $L_\mu-L_\tau$ model. We learned that although including one Higgs triplet can violate the lepton number, the effect is not sufficient  to explain the neutrino data due to the $U(1)_{L_\mu - L_\tau}$ gauge invariance. It was found that a proper symmetric Majorana mass matrix can be obtained when a pair of vector-like leptons and two singlet scalars, which  carry the $L_\mu-L_\tau$ charges, are introduced. In this model, a specific Frampton-Glashow-Marfatia matrix pattern can be achieved  when  some Yukawa couplings are set to vanish. Using the pattern ${\bf A}_1$, we showed that when the neutrino  data within $1\sigma$ errors and cosmological neutrino bound are satisfied,  the involving Higgs-triplet Yukawa couplings have a hierarchy, i.e., $Y_{\mu\tau} \ll Y_{\mu4} \ll Y_{\tau 5} \ll Y_{45} $, and $Y_{45}$ can be $O(0.1)$. As a result, the effective Majorana neutrino mass is below the upper limit of neutrinoless double-beta decay experiment.  Moreover, when the neutrino data are satisfied, it was found that the model can exhibit  interesting phenomena in flavor and collider physics, such as muon $g-2$, $h\to \mu \tau$, $\tau\to \mu Z' Z'$, and $H^{\pm \pm} \to (W^\pm W^\pm, \mu^\pm \tau^\pm)$ decays, although they are not new findings in this paper. \\

%
%

\noindent{\bf Acknowledgments}

This work was partially supported by the Ministry of Science and Technology of Taiwan,  under grant MOST-103-2112-M-006-004-MY3 (CHC).


\end{document}